\documentclass[useAMS,usenatbib,usegraphicx]{mn2e}
\usepackage[fleqn]{amsmath}
\usepackage{txfonts}
\usepackage[T1]{fontenc}
\usepackage{aecompl}
\usepackage{Journals}
\usepackage{hyperref}

\pdfoutput=1

\newcommand\WISE    {\emph{WISE}}
\newcommand\IRAS    {\emph{IRAS}}
\newcommand\Spitzer {\emph{Spitzer}}
\newcommand\SDSS    {SDSS}
\newcommand\C       {\textsc{Clumpy}}
\newcommand\D       {\textsc{Dusty}}
\newcommand\dif     {\hbox{${\rm d}$}}

\newcommand\tV      {\hbox{$\tau_{\rm V}$}}
\renewcommand\deg   {\hbox{$^\circ$}}
\newcommand\etal    {{et al.~}}
\renewcommand\mp    {\hbox{$\,\pm\,$}}
\newcommand{\ph}[1]{{\hbox{\phantom{#1}}}}
\newcommand{\m}{\scalebox{0.75}[1.0]{\( - \)}}
\newcommand{\p}{\scalebox{0.75}[1.0]{\( + \)}}
\def\about          {\hbox{$\sim$}}
\def\c[#1,#2]{\hbox{${\rm W}#1-{\rm W}#2$}}  
\def\s[#1,#2]{\hbox{${\rm W}#1#2$}}          
\def\w[#1]{\hbox{${\rm W}{#1}$}}             
\def\mic            {\hbox{$\mu{\rm m}$}}

\DeclareMathOperator\erfinv{erf^{--1}}

\defcitealias{WISE2}{Nikutta et al., in preparation}
\defcitealias{CH2003}{CH03}
\defcitealias{vdMarel_Cioni_MC1_2001}{MC1}
\defcitealias{vdMarel_MC2_2001}{MC2}
\newcommand\mci  {\citetalias{vdMarel_Cioni_MC1_2001}}
\newcommand\mcii {\citetalias{vdMarel_MC2_2001}}

\hypersetup{
  bookmarks=false,         
  bookmarksopen=false,     
  pdffitwindow=true,       
  pdftitle={The meaning of WISE colours -- I. The Galaxy and its satellites},    
  pdfauthor={Nikutta et al.},   
  colorlinks=true,         
  linkcolor=blue,          
  citecolor=blue,          
  urlcolor=blue,           
  linktocpage=true
}

\title[WISE colours -- I. Galaxy and satellites]{The meaning of
  \WISE\ colours -- I. The Galaxy and its satellites}

\author[Nikutta \etal]{
\parbox{\textwidth}{
  \begin{flushleft}
    Robert~Nikutta,$^{1,2}\thanks{E-mail: \href{mailto:robert.nikutta@gmail.com}{robert.nikutta@gmail.com}}$
    Nicholas~Hunt-Walker,$^{3}$
    Maia~Nenkova,$^{4}$
    \v{Z}eljko~Ivezi\'{c}$^{3}$ and
    Moshe~Elitzur$^{2}$
  \end{flushleft}
}\vspace{0.4cm}\\
\parbox{\textwidth}{
  $^1$Departamento de Ciencias F\'isicas, Universidad Andr\'{e}s Bello,
      Av. Rep\'{u}blica 252, Santiago, Chile\\
  $^2$Department of Physics \& Astronomy, University of Kentucky,
     Lexington, KY 40506, USA\\
  $^3$Astronomy Department, University of Washington, Box 351580,
     Seattle, WA 98195-1580, USA\\
  $^4$School of English and Liberal Studies, Seneca College, Toronto,
     ON, M2J 2X5, Canada\\
}
}

\date{Version of \today}
\pagerange{\pageref{firstpage}--\pageref{lastpage}} \pubyear{2014}

\begin{document}
\label{firstpage}

\maketitle

\begin{abstract}
  Through matches with the Sloan Digital Sky Survey (SDSS) catalogue
  we identify the location of various families of astronomical objects
  in \WISE\ colour space. We identify reliable indicators that
  separate Galactic/local from extragalactic sources and concentrate
  here on the objects in our Galaxy and its closest satellites. We
  develop colour and magnitude criteria that are based only on \WISE\
  data to select asymptotic giant branch (AGB) stars with
  circumstellar dust shells, and separate them into O-rich and C-rich
  classes. With these criteria we produce an all-sky map for the count
  ratio of the two populations. The map reveals differences between
  the Galactic disc, the Magellanic Clouds and the Sgr Dwarf
  Spheroidal galaxy, as well as a radial gradient in the Large
  Magellanic Cloud (LMC) disc. We find that the C:O number ratio for
  dusty AGB stars increases with distance from the LMC centre about
  twice as fast as measured for near-IR selected samples of early AGB
  stars. Detailed radiative transfer models show that \WISE\ colours
  are well explained by the emission of centrally heated dusty shells
  where the dust has standard properties of interstellar medium (ISM)
  grains. The segregation of different classes of objects in \WISE\
  colour space arises from differences in properties of the dust
  shells: those around young stellar objects have uniform density
  distributions while in evolved stars they have steep radial
  profiles.
\end{abstract}


\section{Introduction}
\label{sec:intro}

Infrared emission from radiatively heated dust carries important
information about the distribution of material around the heating
source. This information can be deciphered either through detailed
imaging and analysis of the spectral energy distribution (SED) of
individual objects, or by studying the SED properties (such as
colours) of statistically large populations of sources. The pioneering
\emph{Infrared Astronomical Satellite} (\IRAS; launched in 1983)
all-sky survey ushered in the era of modern highly successful
astronomical surveys \citep{IRAS1,IRAS2} . The \IRAS\ survey provided
unprecedented opportunity to classify the infrared properties of about
350\,000 astronomical objects using a homogeneous data set obtained
with a single facility (e.g. \citealt{vdVeen}).  Thanks to sensitivity
improvements of up to three orders of magnitude compared to \IRAS, the
recent \emph{Wide-field Infrared Survey Explorer} (\WISE; launched in
2010) all-sky survey detected about 560 million objects
\citep{Wright+2010}.  \WISE\ represents the next major step in
surveying and understanding of the infrared sky. For example, data
from the \IRAS\ point source catalogue (PSC) showed that certain
Galactic objects tend to cluster in well-defined regions of \IRAS\
colour--colour (CC) diagrams, and the same clustering is expected in
the \WISE\ photometric system but for a significantly larger sample
that probes a much larger volume of the Galaxy. Furthermore, and
perhaps more importantly, \WISE\ data are deep enough for its
catalogues to contain a very large number of extragalactic sources,
and they too are clustered in well-defined regions of \WISE\ CC and
colour-magnitude (CM) diagrams. Although \IRAS, too, detected some
extragalactic objects (for a review see \citealt{Soifer1987}), \WISE\
can be considered a true IR counterpart to modern optical surveys of
extragalactic sources, such as the Sloan Digital Sky Survey (SDSS;
\citealt{SDSS}).

%
\begin{figure}
  \includegraphics[angle=0,width=\hsize]{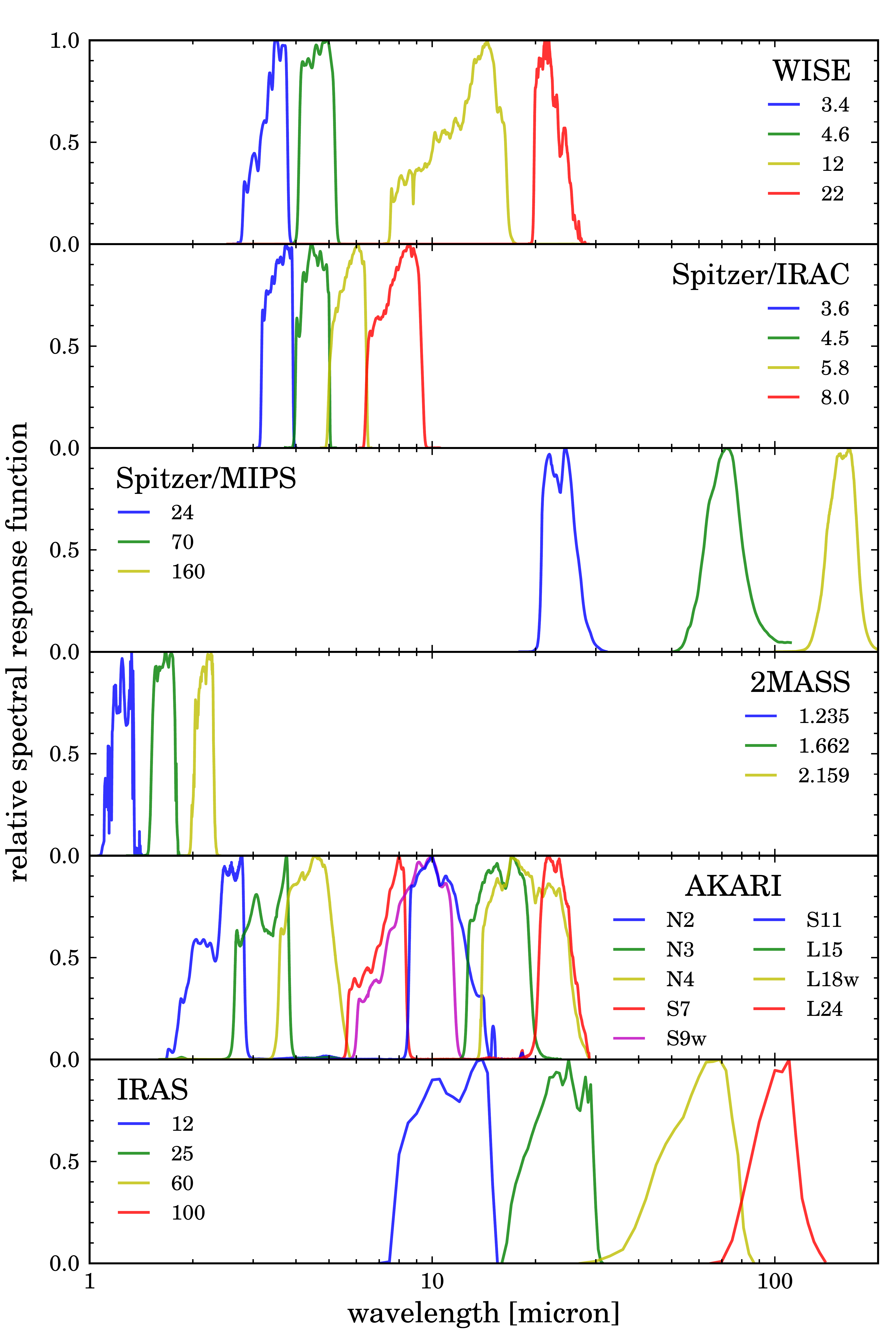}
  \caption{A comparison of bandpass filters for \WISE\ and other major
    infrared surveys. From top to bottom: \WISE\, \Spitzer/\emph{IRAC},
    \Spitzer/\emph{MIPS}, 2MASS, \emph{AKARI}, \IRAS. 2MASS is the only
    ground-based survey. The central filter wavelengths (in micron) or
    the established filter names are given in the legends.}
  \vskip 0.1in
  \label{fig:bands}
\end{figure}
%

\WISE\ mapped the sky at 3.4, 4.6, 12, and 22~\mic\ with an angular
resolution of 6.1, 6.4, 6.5 and 12.0 arcsec,
respectively. Fig.~\ref{fig:bands} shows the \WISE\ bandpasses and
their comparison to other major infrared surveys.  Throughout this
paper we will refer to the filter names as `W1 filter' (and W2, W3,
and W4 filters). The source SED $\lambda F_{\lambda}$ is averaged in
each band by the instrument's transmission function $W_i(\lambda)$,
where $i = 1,2,3,4$. This yields the in-band fluxes $F_i=\int\!
W_i(\lambda) \cdot \lambda F_{\lambda}\, \dif\lambda$. The \WISE\
Catalogue provides magnitudes calibrated in the Vega system using an
empirical fit to the Vega spectrum \citep[see equation~2
in][]{Wright+2010}. Thus the calibrated in-band fluxes are
\begin{equation}
  \label{eq:filterflux}
  f_i = \frac{F_i}{F_i^{\rm Vega}} =
        \frac{\int\! W_i(\lambda) \cdot \lambda F_{\lambda}\, \dif\lambda}
             {\int\! W_i(\lambda) \cdot \lambda F_{\lambda}^{\rm Vega}\, \dif\lambda}.
\end{equation}
`Colour' is the magnitude difference between two measured fluxes of an
astronomical source, e.g., $\c[1,2] = 2.5\log(f_2/f_1)$ for bands 1
and 2. We will also use the short-hand notation \s[1,2] for such
colour.

\WISE\ attained 5$\sigma$ point-source sensitivities better than
0.068, 0.098, 0.86 and 5.4~mJy in the \w[1] through \w[4] bands,
respectively, which correspond to Vega-based magnitudes 16.83, 15.6,
11.32 and 8.0\footnote{See the \WISE\ Explanatory Supplement:
  \url{http://wise2.ipac.caltech.edu/docs/release/allsky/expsup/}\label{foot:explsuppl}}.
These survey sensitivity limits apply in unconfused regions on the
Ecliptic; they improve towards the ecliptic poles due to denser
coverage and a lower zodiacal background. Photometric measurements of
sources brighter than approximately 8.1, 6.7, 3.8 and -0.4~mag (Vega)
are affected by saturation of the detectors. The astrometric precision
for sources brighter than SNR=40 is about 0.2~arcsec.

The \WISE\ All-Sky Release comprises all data taken during the full
cryogenic mission phase (2010 January 7 to 2010 August 6). It includes
an Atlas of 18\,240 images, a Source Catalogue with positional and
photometric data for over 563 million objects, and also an Explanatory
Supplement. Unlike \IRAS\ catalogues which reported fluxes in Jansky,
\WISE\ measurements are reported on the Vega magnitude scale. Although
this difference makes a direct comparison with some \IRAS-based
studies more difficult, we follow already published work based on
\WISE\ data and use unaltered catalogued Vega-based magnitudes. A more
detailed discussion can be found in \cite{Wright+2010} and at the
\WISE data release website\footnote{See:
  \url{http://wise2.ipac.caltech.edu/docs/release/allsky/}}.
%
\begin{figure*}
  \includegraphics[width=0.75\textwidth]{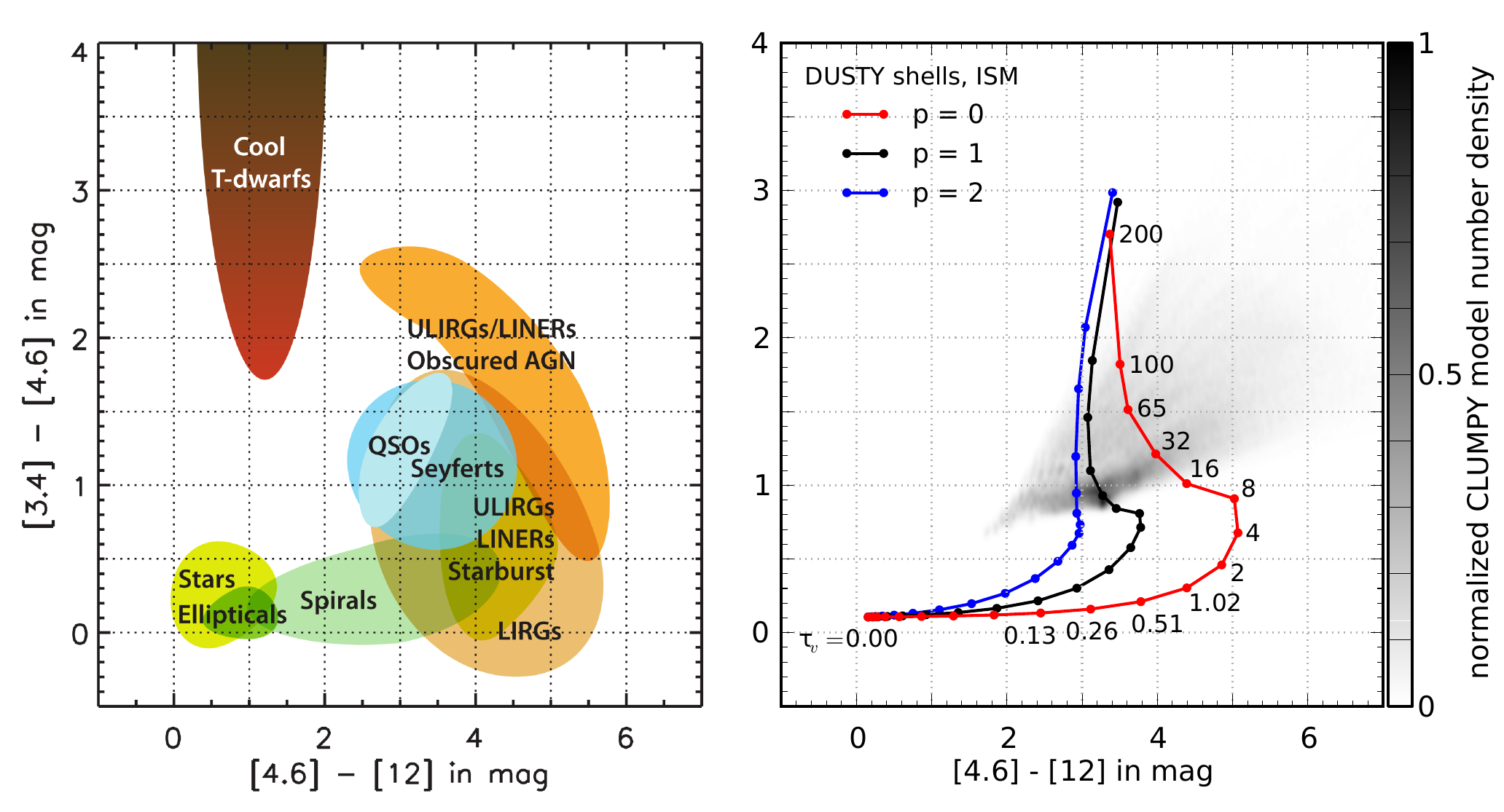}
  \caption{Qualitative comparison of \WISE\ colours with models. Left:
    schematic distribution of \WISE\ colours for various families of
    astronomical sources. This is fig.~12 from \citet{Wright+2010},
    reproduced with permission of the authors and AAS. Right: sample
    model predictions for standard ISM dust. Solid lines show colour
    tracks for IR emission from dusty spherical shells that are
    centrally heated by a 5000~K blackbody point source to a dust
    temperature of 1000~K on the shell inner boundary. The shell
    radial density profile is $r^{-p}$, with $p$ = 0, 1 and 2 shown in
    red, black and blue tracks, respectively. Positions along each
    track correspond to the shell optical depth at visual, as marked
    on the red track. Results of AGN clumpy torus models (see
    \citetalias{WISE2}) are shown as a grey-scale number density
    distribution normalized to its peak value.}
  \label{fig:compare-to-wise}
\end{figure*}

\cite{Wright+2010} have demonstrated that \WISE\ detected all the main
families of extragalactic sources, including star-forming (SF)
galaxies, galaxies with active nuclei (AGN) and quasars (QSO). Their
positions in a representative \WISE\ CC diagram are shown
schematically in the left panel of Fig.~\ref{fig:compare-to-wise}, and
the challenge to theory is to come up with an explanation as to why
different types of objects segregate the way they do. Similar
structuring has been observed in \IRAS-based CC diagrams; its details
were explained by \citet{IE2000}. The colours of a `naked' radiation
source correspond to a point in colour space; since \WISE\ colours are
normalized to Vega, colours of blackbody emission are close to
0. Embedding the source in a dust shell, the radiation is reprocessed
to longer wavelengths and the colours are shifted to the red. The
larger is the dust optical thickness, the redder are the
colours. Therefore, increasing the shell optical depth while keeping
all its other properties constant produces a track in CC diagrams with
distance from the origin along the track increasing with optical
depth. The right panel of Fig.~\ref{fig:compare-to-wise} shows for
illustration three tracks produced by spherical ISM-dust shells with
power-law radial density distributions of varying steepness. The
tracks for uniform density and for $1/r$ density falloff nicely
outline the boundaries of the region occupied by stars and
extragalactic sources in the \WISE\ CC diagram. Similarly, the \WISE\
colours produced by clumpy torus models
\citep{Nenkova+2008a,Nenkova+2008b}, shown as grey-scale number
density distribution, match the region of QSO and Seyferts. Our aim is
to present in this paper a more detailed analysis of \WISE\ colours of
sources in the Galaxy and its satellites, and in a companion paper for
AGN and quasars (\citetalias{WISE2}).

We begin this first paper by identifying the \WISE\ colours of
families of astronomical sources through matching with the \SDSS\
catalogue. The matching criteria and their results are described in
Section~\ref{sec:sdss}. In particular, almost 40 million \WISE\
sources are matched with Galactic stars, and in
Section~\ref{sec:bright-objects} we concentrate on the nature of these
stellar sources. We proceed to calculate model colour tracks for the
\WISE\ photometric system which can be used to interpret the observed
clustering of different stellar populations in the CC diagrams. The
models and their comparison to \WISE\ data are described in detail in
Section~\ref{sec:models}. Our results are summarized and discussed in
Section~\ref{sec:summary}.

\section{\SDSS\ matches for \WISE\ objects}
\label{sec:sdss}

%
\begin{figure}
  \centering
  \includegraphics[width=1.0\hsize]{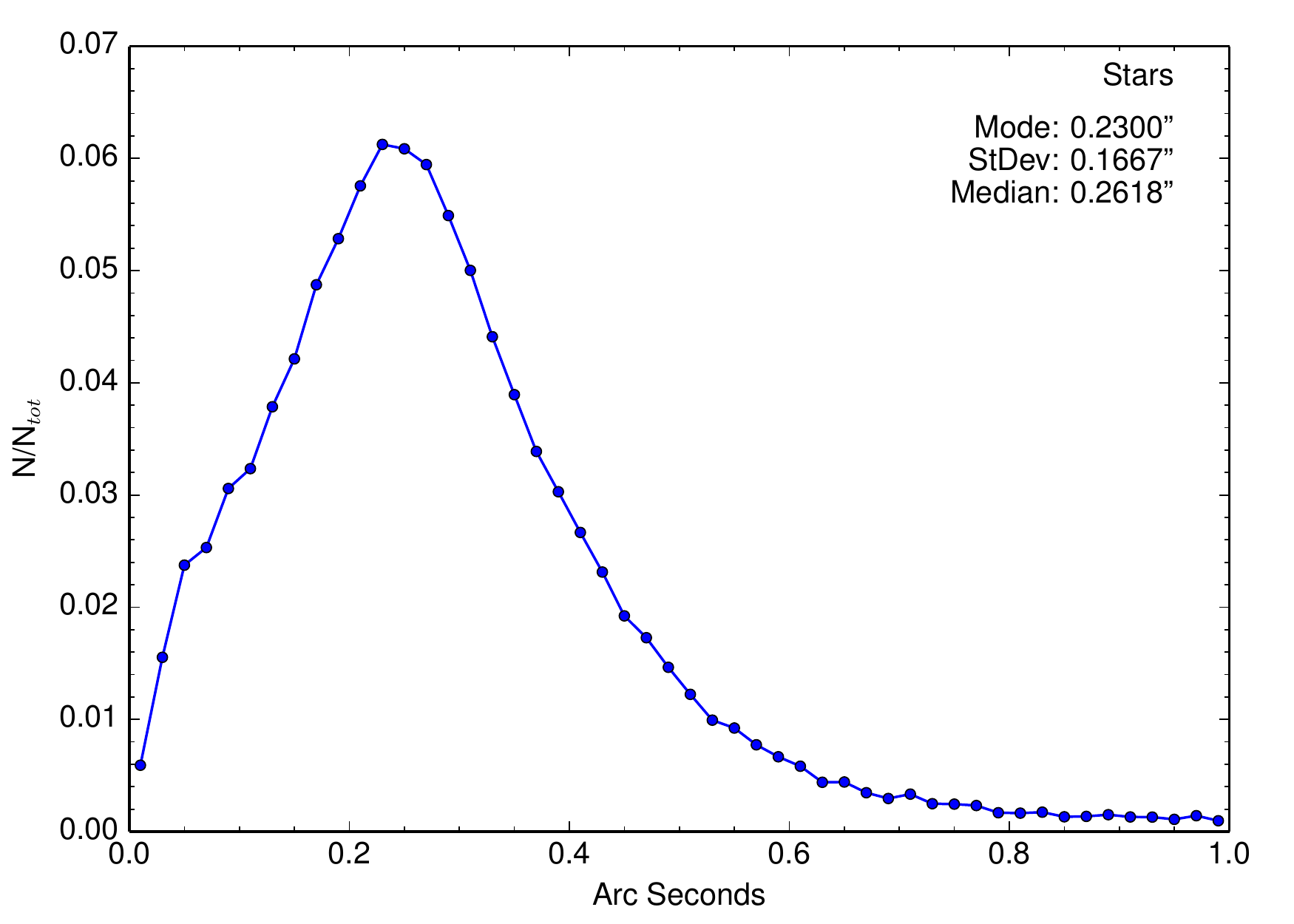}
  \caption{Distribution of matching radii (in arcsec) between \WISE\
    and \SDSS\ for our sample of stars. Based on this figure we adopt
    a matching radius of 1~arcsec in this paper.}
  \label{fig:matching}
\end{figure}
%

\begin{figure*}
  \includegraphics[width=0.8\textwidth]{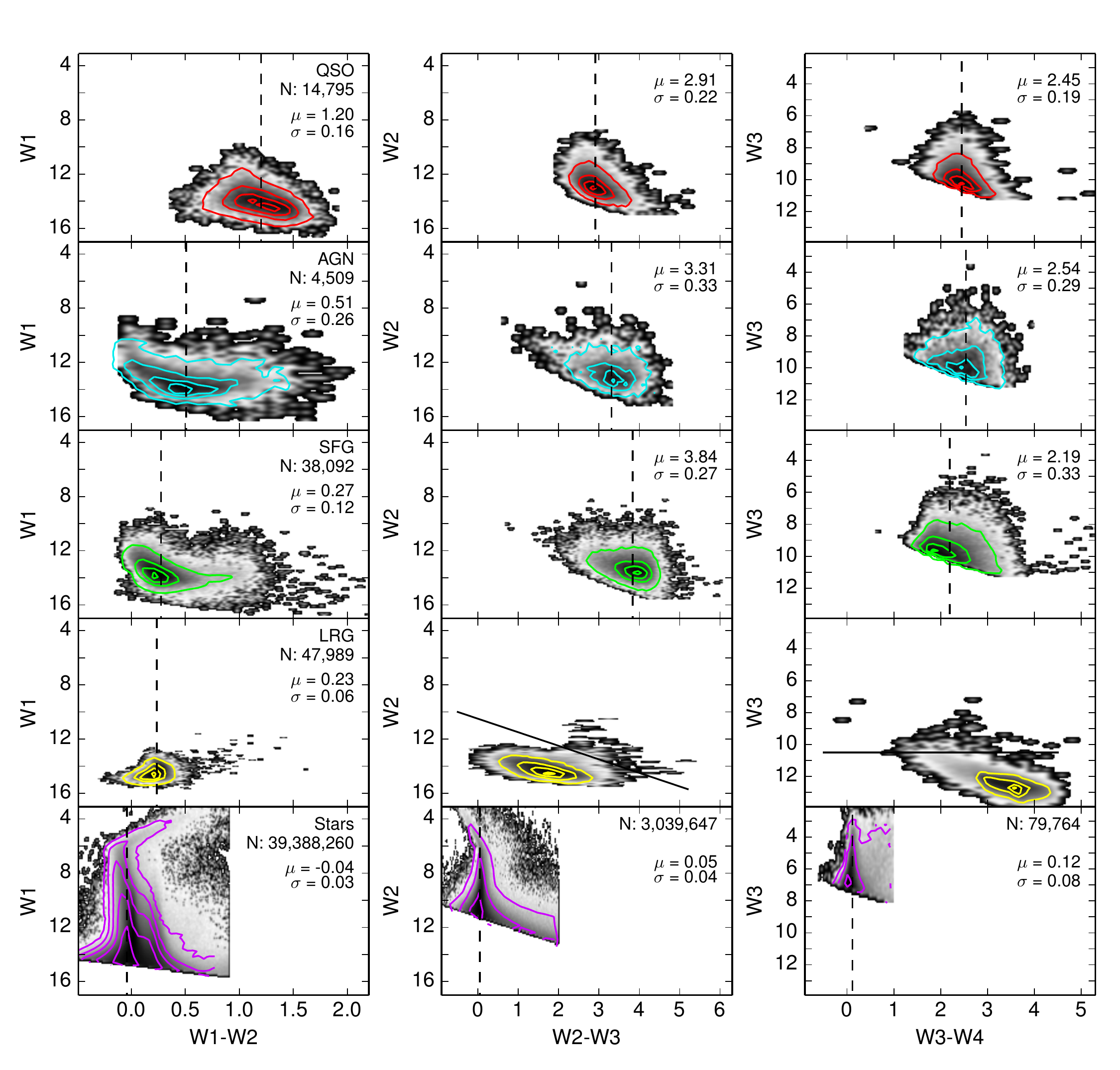}
  \caption{Representative \WISE\ CM diagrams for the five major
    families of objects (rows of panels) classified using \SDSS\
    data. The source distribution is rendered using a logarithmic
    grey-scale map and contours showing 5, 30, 60 and 90 per cent of
    peak density for each non-stellar sample. For the stars (lower
    panels), the left sample shows contours at 35, 50, 65, 80 and 95
    per cent, the middle one at 55, 75 and 95 per cent, and the right
    one at 45, 70 and 95 per cent. The number of plotted objects is
    listed in the top-right corner of each left panel, and it is the
    same for the entire row unless printed otherwise. The solid lines
    in the two right-most panels for LRGs show the 10$\sigma$ limit
    for the \w[3] band: the colours of objects with fainter magnitudes
    are not reliable. Vertical dashed lines in all other panels show
    median values $\mu$ for each distribution, which is indicated
    together with the corresponding interquartile range $\sigma$.}
  \label{fig:color-mag-scatter}
\end{figure*}

To enable a quantitative identification of source populations and
subsequent comparison with our models, we use a sub-sample of \WISE\
sources with \SDSS-based identifications for five types of object:
quasars, galaxies classified as AGN, SF and luminous red galaxies
(LRG; essentially elliptical galaxies), and stars (optically
unresolved objects not in the quasar sample). We first employ
positional matching within a 1.0 arcsec radius of the \WISE\ All-sky
Data Release Catalogue, using procedures described in \cite{Obric} and
\cite{Covey}: we match with \SDSS\ Data Release 7 Catalogue for
samples with spectroscopy, and Data Release 8 for unresolved sources
in imaging data. The 1~arcsec cutoff matching radius is well
justified. Fig.~\ref{fig:matching} shows the distribution of matching
radii between \WISE\ and \SDSS\ for our sample of stars: the median
and standard deviation are $0.26\pm0.17$~arcsec. For a sample of LRGs
and star forming galaxies (not shown) it is
$0.52\pm0.26$~arcsec. After positional matching, we enforce quality
cuts for \WISE\ photometry, as described further
below. Fig.~\ref{fig:color-mag-scatter} shows three representative
\WISE\ CM diagrams for the five sub-samples selected as follows.

%
%

\begin{itemize}
\item {\it Quasars:} starting with 98\,047 sources from the \SDSS\
  catalogue of quasars \citep{Schneider2010} that have a \WISE\ source
  closer than 1 arcsec, we further require that all four \WISE\
  magnitudes are brighter than 5$\sigma$ limits and have a reliably
  measured uncertainty, yielding a sample of 14\,795 quasars.
\item {\it Star forming and AGN galaxies:} Galaxies with \SDSS\
  spectra are selected from the NYU VAGC Catalogue\footnote{Available
    from \url{http://sdss.physics.nyu.edu/vagc/}} \citep{vagc}.
  Galaxies with spectral emission lines are further classified as SF
  and AGN galaxies using the Baldwin-Phillips-Terlevich diagram
  \citep{BPT}. There are 12\,379 AGN and 180\,329 SF galaxies within 1
  arcsec of a \WISE\ source. The cuts for their \WISE\ photometry are
  the same as for quasars, yielding samples of 4509 AGN and 38\,092 SF
  galaxies.
\item {\it LRGs:} luminous red galaxies are adopted from a catalogue
  constructed for cosmological studies by \cite{Kazin2010}. There are
  66\,093 galaxies from that catalogue within 1 arcsec of a \WISE\
  source.  Requiring at least a 5$\sigma$ detection in each of the
  four \WISE\ bands leaves only 70 sources because of faint \w[3] and
  \w[4] magnitudes. As a result, for LRGs we enforce the 5$\sigma$
  cuts only in the \w[1] and \w[2] bands. This yields a sample of
  47\,989 galaxies, whose \c[2,3] and \c[3,4] colours are, of course,
  {\it not reliable}. To emphasize this point, the 10$\sigma$ limits
  for the \w[3] and \w[4] bands are shown as solid lines in the two
  corresponding panels in Fig.~\ref{fig:color-mag-scatter}.
\item {\it Stars:} \SDSS\ Data Release 8 contains 119 million
  unresolved objects within 1 arcsec of a \WISE\ source. To ensure
  high reliability of \SDSS\ star/galaxy separation we limit the
  sample to $r<20$. The majority of these stars are on the main
  sequence, and sample both disc and halo populations \citep[see][for
  detailed discussion]{IBJ2012}.  Similarly to the case of LRGs, we
  apply cuts sequentially.  We first require \w[1] $<$ 15.8 and \w[2]
  $<$ 14.8 (10$\sigma$ limits instead of 5$\sigma$ because the sample
  is already fairly large), and $\s[1,2] < 0.8$ (to avoid
  contamination by faint unrecognized quasars; see below), yielding 39
  million sources (bottom left panel in
  Fig.~\ref{fig:color-mag-scatter}). Applying the 10$\sigma$ cuts also
  to \w[3] and \w[4] bands reduces drastically the sample size,
  yielding three million sources after the $\w[3] < 10.5$ cut (bottom
  centre panel), and $\sim$80\,000 sources when cuts in all four bands
  ($\w[4] < 7.2$) are enforced (bottom right panel).
\end{itemize}

The final sample sizes for all object classes and their mean \WISE\
colours are listed in Table \ref{tab:wisecolors}. The main conclusion
derived from this analysis of \SDSS-\WISE\ matches is that subsamples
of {\it WISE sources with \w[1] $<$ 11 must be dominated by Galactic
  objects}, with the increasing fraction of extragalactic objects at
fainter magnitudes.
%
\begin{table}
  \begin{center}
    \caption{Median and interquartile-based standard deviation,
      \hbox{$\sigma=0.74(q_{75}-q_{25})$}, of \WISE\ colours for the
      five main source classes. For selection details, see
      Section~\ref{sec:sdss}.}\label{tab:wisecolors}
      \begin{tabular}{lrccc}
        \hline
        Class                   & No. selected & \c[1,2]          & \c[2,3]     & \c[3,4] \\
        \hline
        QSO$^{\displaystyle a}$ & 14\,795     & \ph{-}1.20\mp0.16 & 2.91\mp0.22 & 2.45\mp0.19 \\
        AGN                     & 4509      & \ph{-}0.51\mp0.26 & 3.31\mp0.33 & 2.54\mp0.29 \\
        SF                      & 38\,092     & \ph{-}0.27\mp0.12 & 3.84\mp0.27 & 2.19\mp0.33 \\
        LRG                     & 47\,989     & \ph{-}0.23\mp0.06 & n/a         & n/a         \\
        Stars                   & 79\,764      & -0.04\mp0.03      & 0.05\mp0.04 & 0.12\mp0.08 \\
        \hline
      \end{tabular}
  \end{center}
  $^{\displaystyle a}$Without \emph{K}-correction (see \citetalias{WISE2}).\\
\end{table}

\begin{figure}
  \centering
  \includegraphics[width=1.\columnwidth]{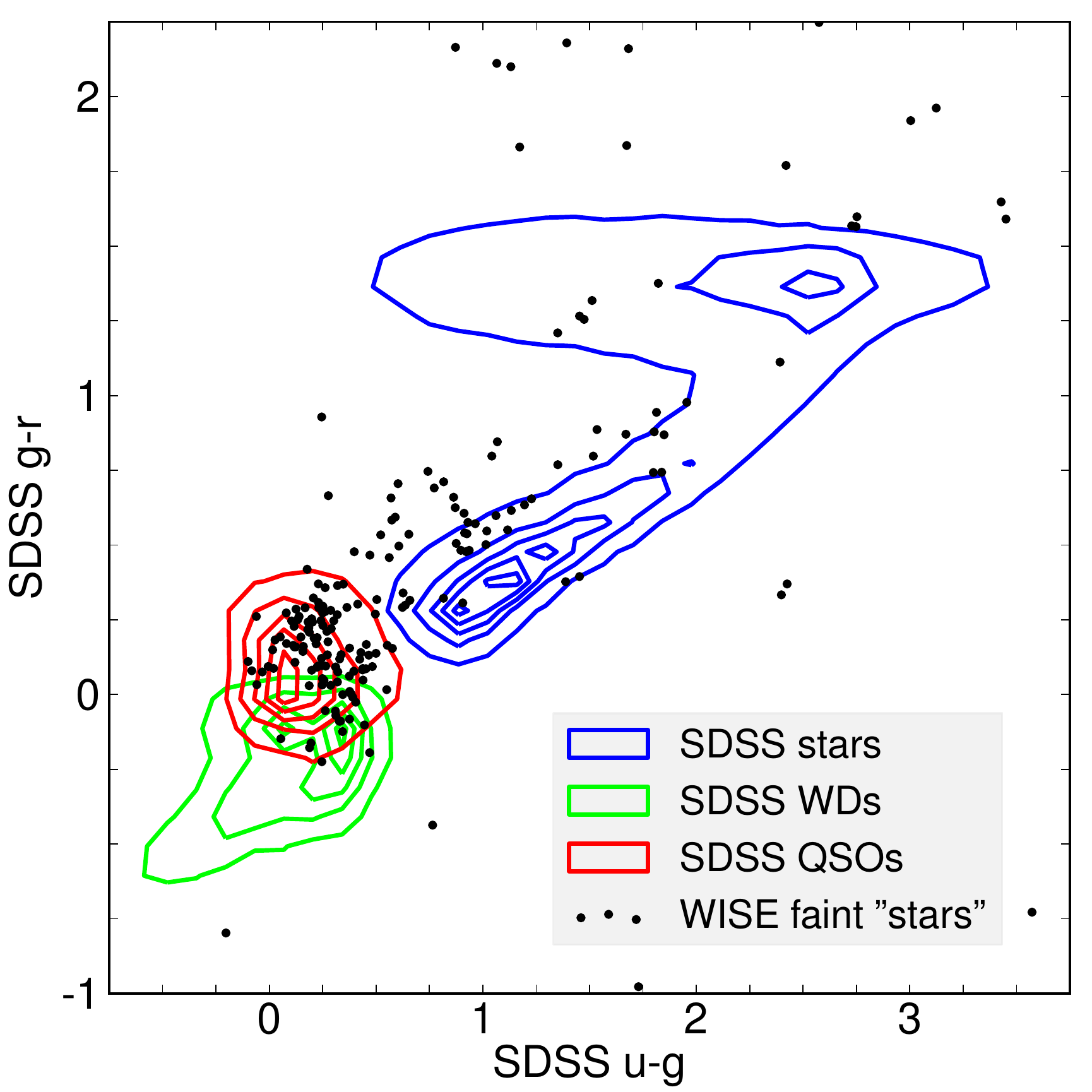}
  \caption{Faint \WISE\ sources in \SDSS\ g-r versus\ u-g CC
    diagram. Objects that have \WISE\ $\c[1,2]>0.8$ and $\w[1] \le 14$
    but are unresolved in \SDSS\ imaging are shown as dots. The
    distributions of true \SDSS\ stars, white dwarfs, and quasars, are
    shown with contours, as indicated in the legend. All contour
    levels are 0.1--0.9 in steps of 0.2. Most of the faint \WISE\
    objects are consistent with quasars that were too faint to be
    listed in the \SDSS\ quasar catalogue.}
  \label{fig:sdss-wise-faint-stars-qsos}
\end{figure}
%

\subsection{Quasar contamination of the faint stellar sample}

For faint `stars', we required $\s[1,2] < 0.8$ because \WISE\ sources
with $\w[1] \ga 14$ and $\s[1,2] > 0.8$ seem to be dominated by
quasars that were too faint to be listed in the \SDSS\ quasar
catalogue. Strong support for this hypothesis is provided by their
distribution in the \SDSS\ $g-r$ versus $u-g$ CC diagram shown in
Fig.~\ref{fig:sdss-wise-faint-stars-qsos}. Objects that are unresolved
in \SDSS\ imaging data and which have \WISE\ $\c[1,2]>0.8$ and $\w[1]
\le 14$ are shown as dots. All other selection criteria for stars as
defined in Fig.~\ref{fig:color-mag-scatter} were also applied. The
distributions of true \SDSS\ stars, white dwarfs, and quasars, are
shown as blue, green, and red contours, respectively. We separated the
WDs from the \SDSS\ stellar catalogue (940 objects out of a total of
$>430$k) according to fig. 23 in \citet{Ivezic+2007}, requiring that
$u-g < 0.5$ and $g-r < 0$.

Most of our faint \WISE\ objects (218 in total, although only 190 dots
are visible within the limits of the figure) are found in the region
occupied by quasars (see, e.g., fig.~1 in \citealt{Smolcic+2004}). A
smaller fraction is further consistent with anomalously red quasars,
found along and just above the stellar locus (for details, see
\citealt{Richards2003}). Finally, only very few of the faint objects
could be misclassified WDs. White dwarfs, although of similar \SDSS\
$u-g$ colour as QSOs, tend to be significantly bluer in \SDSS\ $g-r$.

\section{Colour distribution of bright objects}
\label{sec:bright-objects}

\begin{figure*}
  \center
  \includegraphics[width=\hsize]{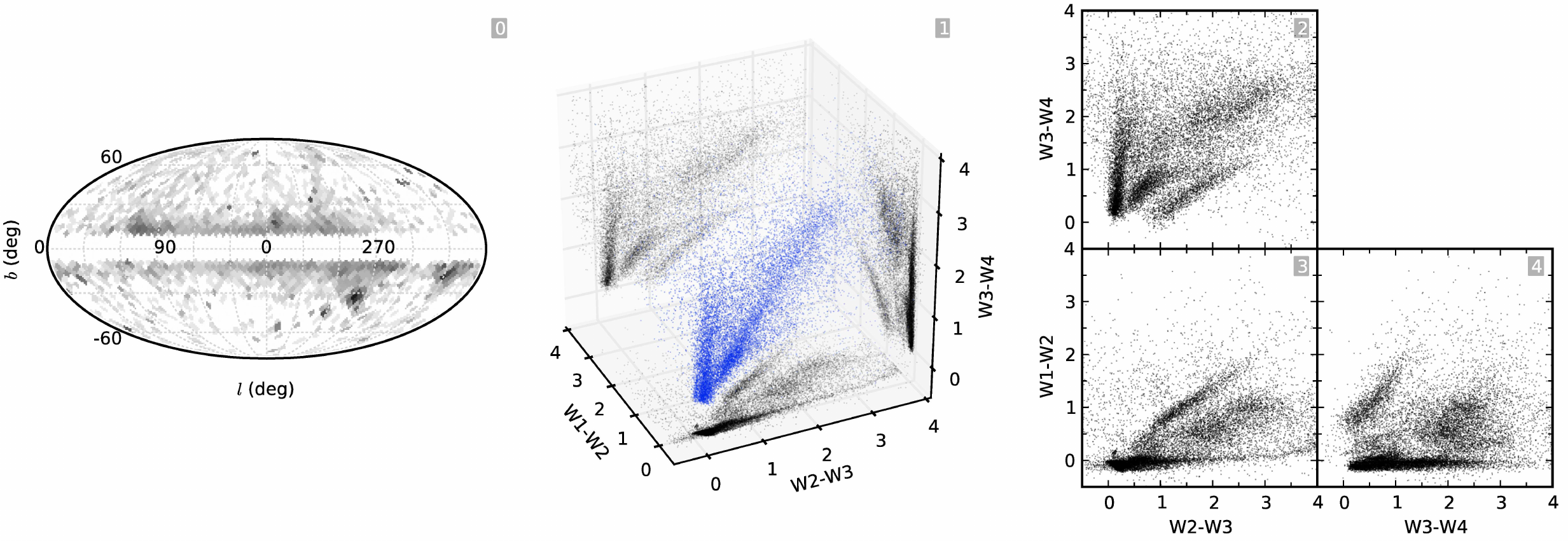}
  \caption{Distribution of 16\,000 bright sources with Galactic
    latitude $|b| > 10\deg$ on the sky and in \WISE\ colour space; see
    text for all selection criteria. Each panel is numbered in its
    upper right corner. (0): All-sky map, shown in a Mollweide
    projection with Galactic coordinates. The logarithmic grey-scale
    corresponds to the number of sources per unit area in the
    sky. (1): 3D distribution (shown in blue) of the three independent
    \WISE\ colours \s[1,2], \s[2,3] and \s[3,4]. The orthogonal
    projections on the corresponding CC planes are shown in black
    scatter diagrams.  (2--4): The three \WISE\ CC diagrams (sides of
    the cube in panel~1). Several blobs and plumes are clearly
    discernible.}
  \label{fig:star-plumes}
\end{figure*}

The analysis of \SDSS-\WISE\ matches shows that \WISE\ sources with
\w[1] $<$ 11 must be dominated by Galactic objects. In this section we
analyse this bright subset in more detail. We start by excluding
objects close to the galactic plane ($|b|<10\deg$) due to unknown
levels of extinction. We then require that the sources have magnitudes
between the 5$\sigma$ sensitivities and the saturation limits in all
four bands. We also ask that the signal-to-noise ratio be greater than
5 in bands \w[1] and \w[2], and greater than 10 in bands \w[3] and
\w[4]. Finally, by imposing $\w[1]<11$ and $\w[2] < 10$ we end up
mostly with stars. We produced an all-sky map of the selected
$\sim$16\,000 bright sources by subdividing the sky into small patches
of constant area on the sphere, about 13.43\,deg$^2$, using the
\textsc{Healpix} tessellation scheme
\citep{HEALPIX}.\footnote{\textsc{Healpix} is available at
  \url{http://healpix.sourceforge.net}. We specifically used the
  \textsc{Python} wrapper \textsc{Healpy}, available at
  \url{https://github.com/healpy/healpy}.} The number of objects in
each \textsc{Healpix} patch was then counted to construct the
logarithmic grey-scale map shown in the leftmost panel (0) of
Fig.~\ref{fig:star-plumes}. This map, shown in Mollweide projection
with Galactic coordinates, reveals the Galactic bulge and the
Magellanic Clouds.

From the four \WISE\ fluxes one can produce 12 different colours, but
only three of them are independent. We choose \s[1,2], \s[2,3] and
\s[3,4] as our three independent \WISE\ colours. Their 3D distribution
for the bright sources is shown in blue in the middle panel (1) of
Fig.~\ref{fig:star-plumes}, with the 2D orthogonal projections shown
as black scatter plots on the sides of the cube; these are the
corresponding CC planes. These three projections are repeated in
Panels~2--4, which show the \WISE\ CC diagrams for Galactic sources;
if Panels~2 and~4 are folded upwards along their common axes with
Panel~3, the cube in Panel~1 is reproduced. We proceed now to discuss
the structure evident in the \WISE\ colour space.

%
%

\subsection{Naked stars}
\label{sec:naked_stars}

The \WISE\ magnitudes are normalized to the stellar spectrum of Vega,
an A-type star. The colours of most stars should thus cluster around
zero in the CC diagrams. Indeed, the highest concentration of objects
in Figs.~\ref{fig:compare-to-wise} and \ref{fig:star-plumes} is found
around zero colours. These are `naked' stars, i.e. stars without
significant circumstellar IR excess. In the case of \IRAS, all filters
sampled the Rayleigh-Jeans tail of every stellar blackbody (see
Fig.~\ref{fig:bands}) so that naked stars corresponded to a single
point in all CC diagrams. In contrast, \WISE\ bands~1 and~2 cover
wavelengths short enough to be close to the peaks of some cool stellar
spectra, therefore deviations from the zero-position in the CC
diagrams can be expected, primarily for stars significantly cooler
than Vega.

\begin{figure}
  \centering
  \includegraphics[angle=0,width=0.9\hsize]{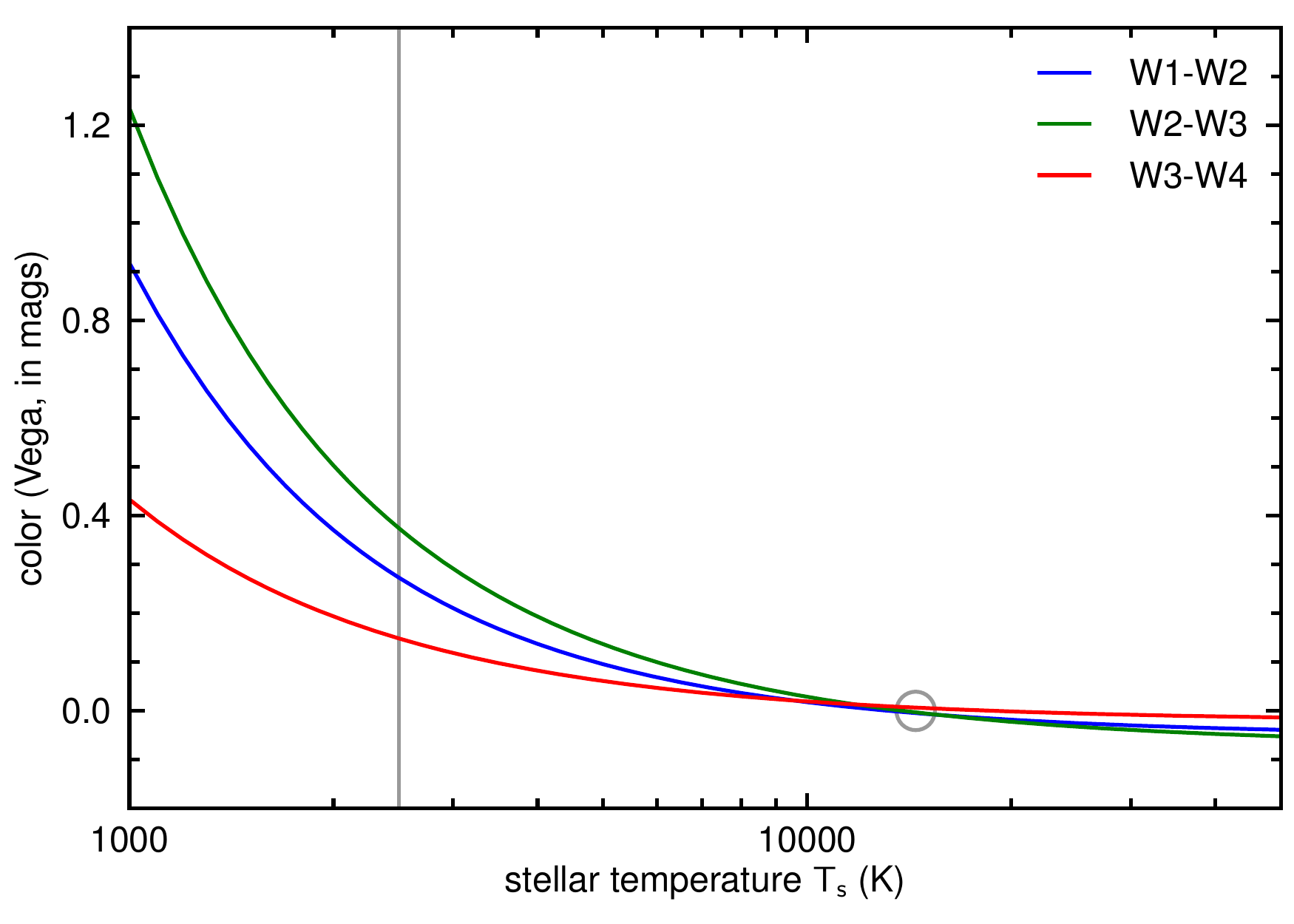}
  \caption{\WISE\ colours of `naked stars', represented by blackbody
    emission with surface temperatures $T_s$. The grey circle
    pinpoints the modified-blackbody empirical fit for Vega, adopted
    by \citet{Wright+2010}. The grey vertical line at $T_s$ = 2500~K
    marks a typical AGB stellar temperature (see
    Sections~\ref{sec:agb} and \ref{sec:agb-analysis}).}
  \label{fig:bbcolors}
\end{figure}

Fig.~\ref{fig:bbcolors} shows the three orthogonal \WISE\ colours
\s[1,2], \s[2,3], and \s[3,4] for naked stars, described by blackbody
emission with a large range of surface temperatures $T_s$; AGB stars
have $T_s\!\approx\!$~2500~K (see Sections~\ref{sec:agb}
and~\ref{sec:agb-analysis}), while O-type stars reach upward of
50\,000~K. Because \cite{Wright+2010} used for Vega a slight
modification of a 14\,454~K blackbody (see their equation~2), to take
care of instrumental/calibration effects, the colours produced by an
unmodified Planck function at that temperature are not precisely 0.
At higher temperatures, all \WISE\ bands fall in the Rayleigh-Jeans
domain and the three colours become constant with a slight ($\sim
-0.05$~mag) deviation from zero. At the cool end the deviations are
larger, reaching \about +0.4~mag for AGB stars. This is the scatter
observed in the stellar blob around the colour origin (0,0,0) in
Fig.~\ref{fig:star-plumes} (Panels~1--4). All deviations larger than
this intrinsic scatter must be due to radiation reprocessing by a
dusty environment.

%
%

\subsection{Dusty envelopes}
\label{sec:envelopes}

The several plumes and blobs that stand out in the \WISE\ colour space
away from the origin reflect reprocessing of stellar radiation by
surrounding dusty envelopes. Such reprocessing redistributes the
energy to longer wavelengths, leading to redder colours. The
discernible structure is reminiscent of the clustering in \IRAS-based
CC diagrams (e.g. \citealt{vdVeen}; \citealt{IE2000}).  In the case of
Galactic \IRAS\ sources, these CC structures correspond to different
classes of stars. \citet{IE2000} found that they were predominantly
young stellar objects (YSO) and AGB stars, and the same is expected
for the Galactic \WISE\ sources. The narrow, elongated plume parallel
to the \s[2,3] axis at \s[1,2]~$\approx$~0 in Panel~3 of
Fig.~\ref{fig:star-plumes} is produced by warm dust emitting
predominantly at short IR wavelengths. Such warm emission requires
heating by a hot central star and can be expected to be YSO signature.
Indeed, \citet{IE2000} found that the locus of \IRAS\ colours of YSOs
could be reproduced with emission from dust shells with uniform
density. Such shells produce the \WISE\ colours outlined in
Fig.~\ref{fig:compare-to-wise} with a red track \hbox{($p$ = 0)},
which passes through this plume. Thus this plume is the \WISE\ colour
locus of YSO sources, which we discuss further in
Section~\ref{sec:yso}. For now we exclude this plume by retaining only
sources with an IR excess in both `warm' colours, \s[1,2] $> 0.2$ and
\s[2,3] $> 0$, concentrating on the other plumes that comprise the AGB
sources found by \WISE.

\subsection{AGB stars}
\label{sec:agb}

The progenitors of asymptotic giant branch (AGB) stars are red giants;
their progeny are planetary nebulae and white dwarfs. According to
photospheric chemical composition, they can be divided into
oxygen-rich (O-rich) and carbon-rich (C-rich) stars. More evolved AGB
stars have dusty winds driven by radiation pressure on the dust grains
in their circumstellar shells \citep[see e.g.:][and references
therein]{EI01, IE10}.  The winds reprocess the stellar radiation,
shifting the spectral shape towards the infrared.  These dusty AGB
stars can be divided into two major sub-groups according to dust
chemical composition: O-rich stars are associated with silicate-rich
dust chemistry and C-rich stars with carbonaceous dust grains.

Dusty AGB stars are observable in our Galaxy and in its close
satellites, mostly the Magellanic Clouds, e.g., by \IRAS\
\citep{vdVeen}, \emph{ISO} \citep{Trams1999}, \emph{MSX}
\citep{Egan2001}, \emph{Akari} \citep{Ita2008}, and \Spitzer\
\citep{Vijh2009,Bolatto2007} [but see also, e.g., \citet{Menzies+2008,
  Whitelock+2009, Boyer+2013, JD2012, Javadi+2013}, for other Local
Group galaxies]. Studies of the Large Magellanic Cloud (LMC) revealed
an overabundance of C-rich objects among its AGB stars, a trend found
both at the near-IR wavelengths of the \emph{DENIS} survey
(\citealt{CH2003}, hereafter \citetalias{CH2003}) and the longer
wavelengths of \IRAS\ (e.g., \citealt{Zijlstra2006}). Observations by
\IRAS\ and 2MASS of AGB stars in our own Galaxy were limited by
interstellar dust extinction and by the sensitivities of these surveys
(e.g., \citealt{JIK2002}, \citealt{NW2000}).  Thanks to the much
improved sensitivity and spatial resolution of \WISE, the counts of
catalogued dusty AGB stars should increase dramatically.  This is
especially expected for the Galaxy, where a model-based estimate of
the total population of dusty AGB stars count is about 200\,000
\citep{JIK2002}.

\subsubsection{WISE-based selection of dusty AGB stars} 

%
\begin{figure*}
  \includegraphics[width=0.8\hsize]{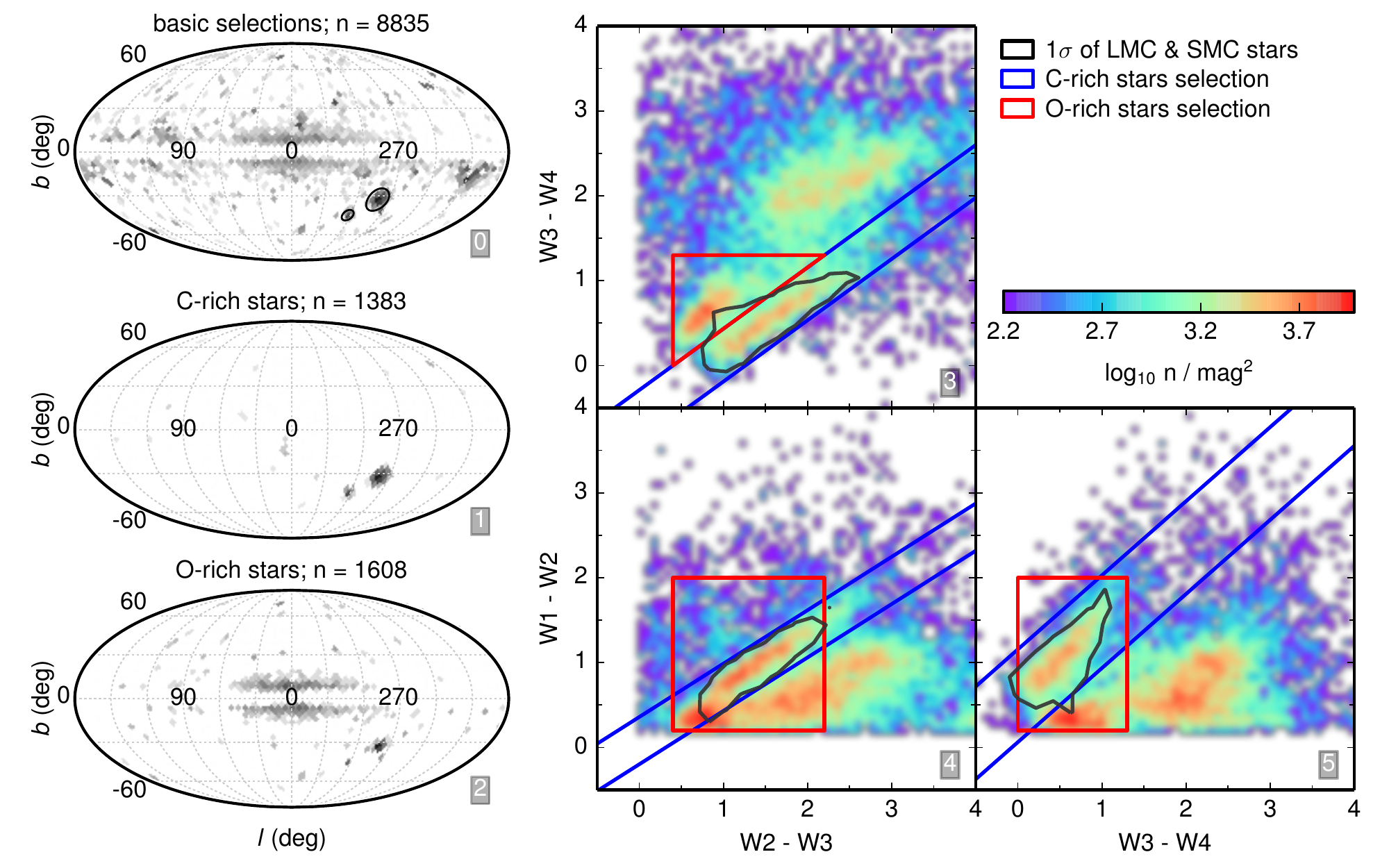}
  \caption{Selection of AGB stars. Each panel is numbered in its lower
    right corner. (0)~All-sky map of bright sources with IR excess in
    Mollweide projection. The Galactic plane with $|b|<6\deg$ has been
    removed. The grey-scale corresponds to the logarithm of the source
    count per unit area on the sky. The LMC and SMC are circled in
    black. (1)~Map of C-rich stars selected in the CC diagrams
    3--5. This sample is dominated by stars in the Magellanic
    Clouds. (2)~Map of the O-rich stars selected by our
    criteria. Stars in the Galactic bulge dominate this
    sample. (3--5)~Three \WISE\ CC distributions of the sources in
    Panel~0. The logarithmic colour scale (common to all CC diagrams)
    shows the number of sources per ${\rm mag}^2$ bin. The black
    contours contain the peak and 68 per cent of all sources within
    the black circles in Panel~0; these are dominated by the C-rich
    stars in the LMC. The regions highlighted with blue and red lines
    delineate our respective selection criteria for C-rich and O-rich
    stars. The cluster of objects around \c[2,3]~$\approx$~2 and
    \c[3,4]~$\approx$~2 contains a mix of Galactic and extragalactic
    sources, but relatively few AGB stars (see Section~\ref{sec:agb}
    for details).}
  \label{fig:agb-selection}
\end{figure*}
%

To remove the YSOs and still maintain a sizeable sample of AGB
sources, first we slightly relaxed the constraint on the Galactic
latitudes and omitted all sources with $|b| < 6\deg$. This excludes
most YSOs, which cluster around the Galactic plane. We arrived at this
choice by testing how well the plumes in Fig. \ref{fig:star-plumes}
still separate visually when allowing latitudes closer and closer to
the Galactic plane. We then removed the narrow, elongated YSO plume as
described above (Section~\ref{sec:envelopes}). Finally, we required that the
stars must also have colours within boundaries determined from known
AGB stars from the SIMBAD data base; a detailed discussion is given in
Appendix \ref{app:agb-selection}. The final sample has 8835
members. Its all-sky distribution is shown in Panel~0 of
Fig.~\ref{fig:agb-selection}, where the logarithmic grey-scale
corresponds to the normalized number of selected stars per unit area
of the sky. Black circles are drawn around the LMC and SMC, with radii
of $\sim$8\deg\ and 4\deg, respectively.

The colours of AGB sources form a substructure of the 3D blue
structure in the \WISE\ colour space, shown in
Fig.~\ref{fig:star-plumes}. For simplicity, in
Fig.~\ref{fig:agb-selection} we show only the 2D projections of this
substructure on the three orthogonal colour planes. Panels~3--5 of
this figure show how the selected stars distribute in \WISE\ CC
diagrams. The black contours contain 68 per cent of all the stars
within the two black circles drawn around the LMC and SMC in Panel~0,
as well as the peaks of their CC distributions. Since dusty AGB stars
in the Magellanic Clouds are dominated by C-rich stars, the plumes
outlined with these black contours mark the locus of C-rich stars in
the \WISE\ colour space. To quantitatively improve the selection of
C-rich stars, we define two slab-shaped regions shown with blue lines
in Panels~3 and~4; the corresponding blue lines in Panel~5 are a
consequence of these definitions. In the 3D colour space, the
intersections of these slabs define a parallelepiped within which the
selected C-rich stars fall. The centre line of this elongated volume
follows roughly the greatest density of the C-rich star plume (see
Appendix \ref{app:agb-selection} for the mathematical details).

Two structures remain. From cross-checks with SIMBAD, the plume
outlined with red is dominated by O-rich AGB stars. The separation
between the O-rich and C-rich plumes is cleanest in the
\s[2,3]--\s[3,4] diagram (Panel~3), where the red triangle is the
locus of O-rich AGB stars. The slanted edge of our selection aligns
with the C-rich star selection boundary. The edge in \s[3,4] is tuned
to exclude additional objects (dominated by quasars, see below). The
\s[2,3] edge is positioned to maximize the completeness of our O-rich
star selection without significant overlap with stars which have very
little IR excess, i.e. which might not be AGB stars with circumstellar
dust shells.  In the two other panels (4 and 5) the separation is not
as clean, thus we use for selection criteria the simple rectangular
areas outlined in red.

The other structure, around \s[1,2] $\approx 0.5$, \s[2,3] $\approx
2.7$, and \s[3,4] $\approx 2.2$, is disconnected from the origin of
the \WISE\ colour space. Its objects are a mix of Galactic and
extragalactic sources outside the selection boundaries for O-rich and
C-rich stars. Cross-checks with SIMBAD show that the blue end of this
plume in Panel~3 comprises purely Galactic sources, mostly YSO,
variable stars, emission-line stars and only $<$10 per cent AGB
candidates. The YSOs most likely contribute to the \w[3] band through
their polycyclic aromatic hydrocarbon (PAH) emission. Thanks to its
especially broad filter (\mbox{$\sim$ 7--18~\mic}; see
Fig.~\ref{fig:bands}), \WISE\ band 3 covers the prominent PAH features
at 7.7, 8.6, and 11.3~\mic, indicative of star formation
\citep[e.g.][]{Peeters+2004}. Such non-thermal, fluorescent emission
is outside the scope of our study. Towards the red end of this plume
we find increasingly more and more extragalactic sources (red QSOs,
Seyferts, etc.) that have slipped through our selections. Thus this
plume is a diverse mix of objects and will be ignored hereafter.

To summarize, we have demonstrated in this section that it is possible
to use \WISE\ photometry to select fairly clean samples of O-rich and
C-rich AGB stars.  We note that our results for the distribution of
C-rich stars differ significantly from those in a recent study by
\cite{TuWang2013}. We believe that their results were affected by
saturation effects in \WISE\ photometry, as discussed in detail in
Appendix \ref{app:agb-selection}.

\subsubsection{Near-IR colours of WISE-selected AGB stars}
\label{sec:AGBnearIR}

The \WISE\-based selection of AGB stars described in the preceding
section is sensitive to the presence of circumstellar dust emission.
The success of this selection, as well as the ability of \WISE\
colours to distinguish envelopes with carbonaceous dust from those
with silicate dust, can be verified using near-IR photometry. Based on
published work (e.g., with 2MASS and \emph{DENIS} surveys;
\citealt{NW2000}, \citetalias{CH2003}), AGB stars with carbonaceous
dust are expected to be redder than $J-K_s \sim 2$, while stars with
silicate dust, and early AGB stars without much dust, have bluer
$J-K_s$ colours.

\begin{figure}
  \includegraphics[angle=0,width=1.\hsize]{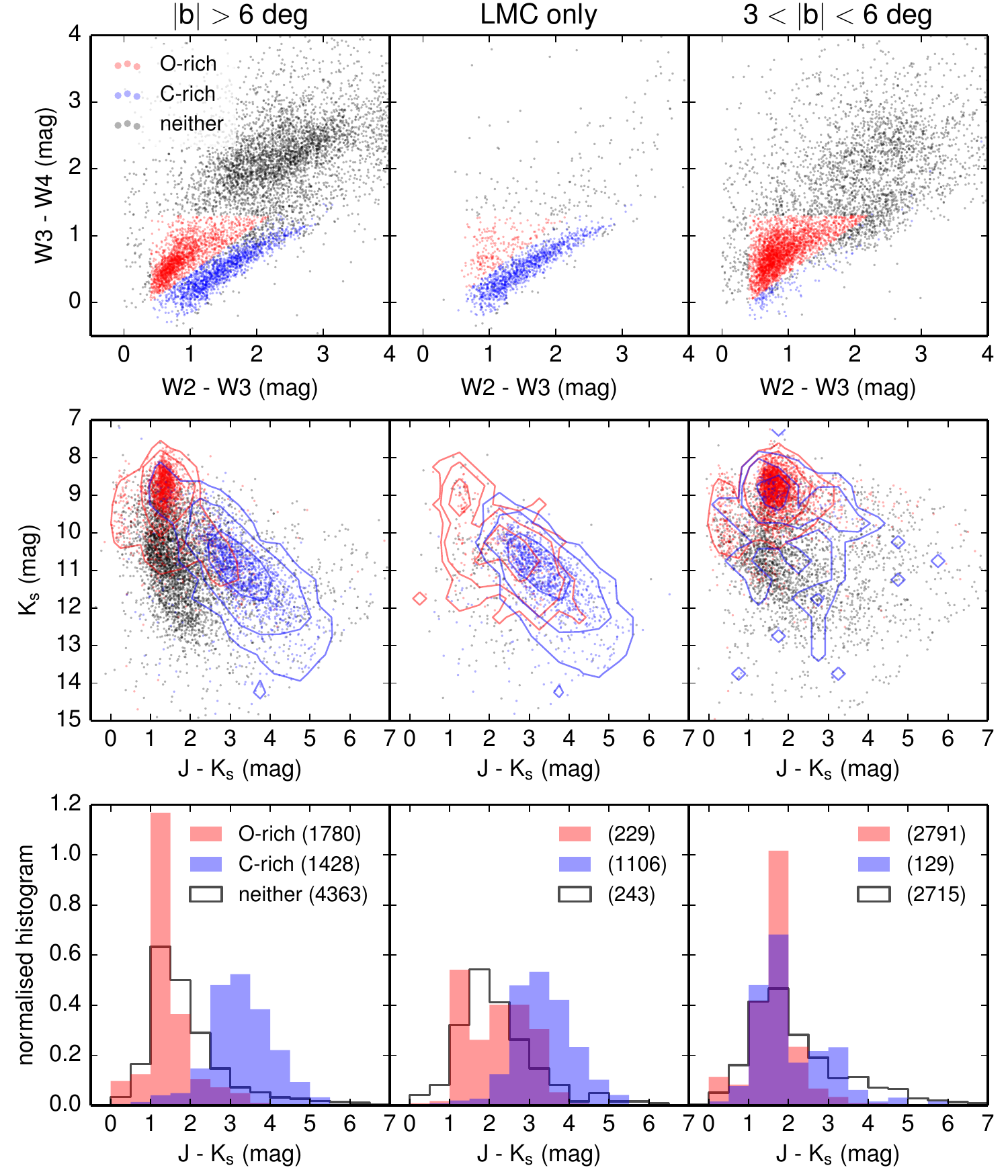}
  \caption{Analysis of near-IR J-K colour distribution for
    WISE-selected stars with circumstellar dust shells. The colour and
    magnitude selection criteria are as in Fig. \ref{fig:co-ratio},
    but additionally the selected stars must have valid measurements
    in 2MASS. The three columns correspond to three different
    positionally-selected subsamples, as indicated at the top. The
    majority of C-rich stars are found in the LMC, and most have
    $J-K_s>2$, consistent with previous work and with models. O-rich
    stars with silicate dust shells predominantly have $J-K_s <
    2$. The O-rich sample from the LMC (middle column) and the C-rich
    sample from low Galactic latitudes (right column) appear
    contaminated by much more numerous samples of C-rich and O-rich
    stars, respectively.}
  \label{fig:2mass}
\end{figure}

Fig.~\ref{fig:2mass} shows the $K_s$ versus $J-K_s$ CM diagrams
constructed with 2MASS photometry for three subsamples of
\WISE\-selected AGB stars. These subsamples are selected positionally
from low and high Galactic latitudes, and from the LMC region. The
expected difference in the $J-K_s$ colour distributions between
WISE-selected subsamples of stars with carbonaceous and silicate dust
is easily discernible. Nevertheless, when the numbers of the two
populations are very different (LMC is dominated by C stars and the
Galaxy by O stars, see below), the smaller population can be
significantly dominated by the larger one.
%
%

The \WISE\-selected AGB stars are biased in favour of dusty AGB stars,
while samples selected using near-IR photometry (and sometimes visual
photometry) are strongly biased against dusty AGB stars. For example,
the well-studied sample selected using visual and near-IR photometry
from \citet{Cioni2009} does not include any stars with $J-K_s>2.3$!
This selection is controlled by direct photospheric emission while
\WISE\ fluxes are dominated by re-radiation from dusty shells. These
different selection criteria make data analysis and interpretation
more difficult, but they are also advantageous because of implicit
constraints on age and metallicity distributions of the underlying
populations, as we discuss below.

\subsubsection{The spatial variation of the C:O star count ratio}
\label{sec:co-ratio}

With the colour-selection criteria for C-rich and O-rich AGB stars
established, Panels~1 and 2 of Fig.~\ref{fig:agb-selection} show the
distributions of these two families in the sky. The C-rich star sample
is dominated by stars from the Magellanic Clouds, while O-rich stars
are found predominantly towards the Galactic bulge and plane.  This
behaviour of dusty AGB stars is also seen for samples selected using
near-IR photometry (recall that the latter are biased in favour of
early AGB stars without thick dusty envelopes): while the Galaxy has
more O-rich stars, the LMC and SMC are dominated by C-rich stars
(\citetalias{CH2003}). This difference is likely due to higher
metallicity in the Galactic disc than in LMC/SMC \citep{BD2011}.

\begin{figure*}
  \includegraphics[width=0.9\hsize]{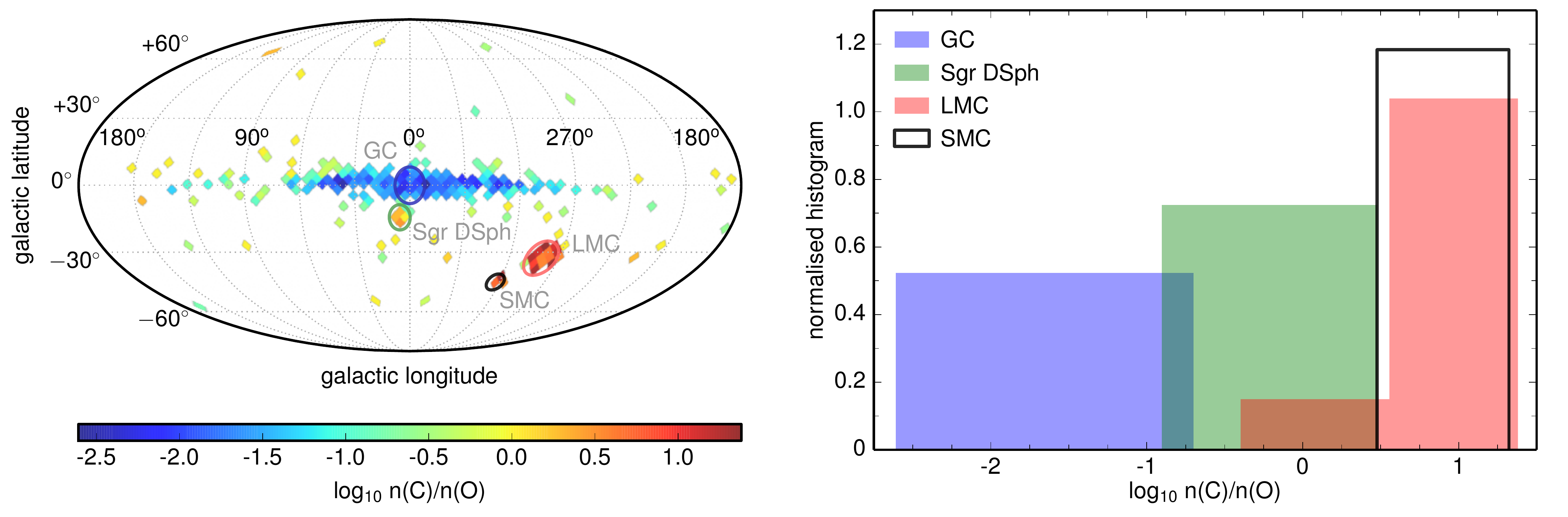}
  \caption{Left: all-sky map of the local C-to-O-rich star number
    ratio, in Mollweide projection. The Galactic Centre is at
    (0,0). This map is a ratio of Panels~1 and 2 in
    Fig. \ref{fig:agb-selection}, but without the $|b|>6\deg$
    criterion. The colour scale shows the logarithm of the ratio per
    13.43~deg$^2$ coordinate bin (defined by the \textsc{Healpix}
    tessellation). The locations of the Galactic Centre and three
    Milky Way satellites are indicated with (projected)
    circles. Right: Distribution of ${\rm \log_{10} n(C)/n(O)}$ in
    each circular area from the left panel. The histograms were
    obtained using Bayesian Blocks (see text). See
    Table~\ref{tab:coratio} for statistics.}
  \label{fig:co-ratio}
\end{figure*}

The all-sky coverage and large size of the \WISE\ sample of AGB stars
allow us to compute a projected map of the C-to-O-rich star number
ratio as a function of sky coordinates. The left panel of
Fig.~\ref{fig:co-ratio} shows in a colour map the variation of this
ratio in the Galaxy and its immediate vicinity. The sky was divided
into small 13.43\,deg$^2$ patches (see
Section~\ref{sec:bright-objects}), and the ratio was determined per
patch. For this map we have dropped the previous requirement to
exclude objects close to the Galactic plane, and selected all stars
based on the colour and magnitude criteria defined previously. A
strong gradient in the ratio, from the lowest values in the Galactic
plane to the highest C-rich star concentration in the Magellanic
Clouds, is clearly visible. The dynamical range spans almost four
orders of magnitude.

Several regions stand out and are marked with (projected) circles: the
Galactic Centre, and the LMC and SMC. Quite surprisingly, the
Sagittarius Dwarf Spherical Galaxy (Sgr DSph) is also clearly
discernible. The coordinates of these four regions are taken from the
NASA/IPAC Extragalactic Database (NED) and are listed in
Table~\ref{tab:coratio} (lines~1 and 2). For the outlined circles we
adopted radii 1.5 times the reported optical semi-major axes (line
3). For the Galactic Centre region we chose a radius identical to the
LMC.

The right panel of Fig. \ref{fig:co-ratio} shows histograms of the C:O
ratio for each sub-region, measured among all non-empty patches in
that region. The histograms were obtained using the ``Bayesian
Blocks'' technique devised by \citet{BayesianBlocks2013}. This method
maximizes Bayesian evidence for a model under the assumption that data
can be described by a piece-wise step function with arbitrarily
located knots (analogous to a histogram with variable bin width). We
used the implementation provided in the
\texttt{plotting.hist\_tools.bayesian\_blocks} function from the
\textsc{astroML} package \citep{astroML,astroMLText}. The C:O-rich star
ratio for Sgr DSph is about 1.5 orders of magnitude higher than in the
Galactic plane, though less than that of the Magellanic Clouds.  Table
\ref{tab:coratio} lists the means (line~8) and ranges (lines~9 \& 10)
of the histograms in Fig. \ref{fig:co-ratio}, together with the values
measured for the full sky. Line 7 also reports the C:O ratios for the
entire sample of stars in each region, without regard for their
spatial distribution within that region. In the LMC, this value
deviates significantly from the mean of all small in-region patches
(line~8).\footnote{A similar deviation is seen in Sgr DSph, but the
  number of contributing patches might be too small for meaningful
  conclusions.} The conclusion is that the C:O ratio in the LMC is not
spatially uniform.

\subsubsection{The C:O star count ratio in LMC}
\label{sec:coratiolmc}

\begin{figure}
  \centering
  \includegraphics[width=\hsize]{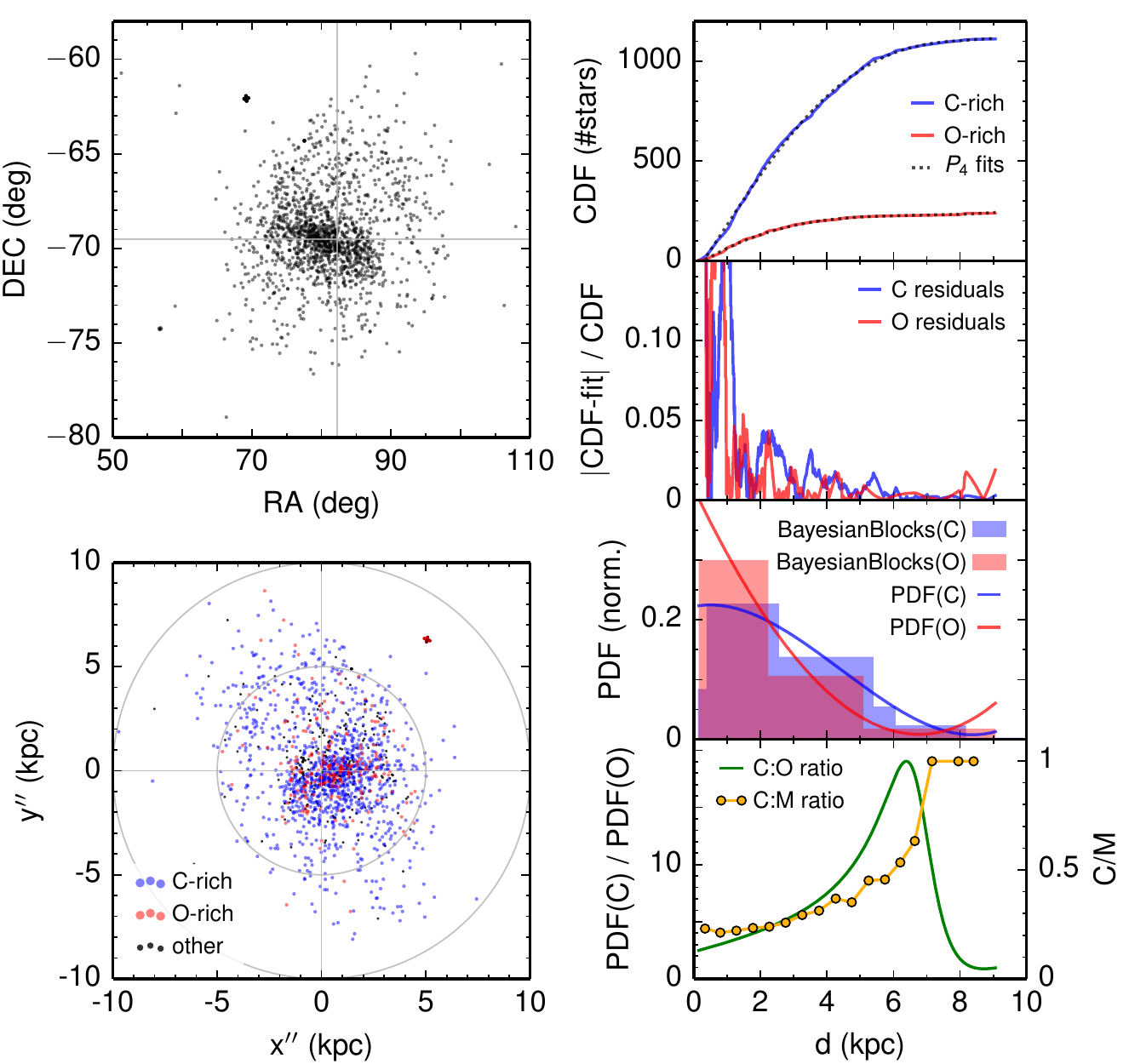}
  \caption{C:O number ratio gradient in the LMC. Left top: stars in
    the LMC in equatorial coordinates. Left bottom: de-projected map
    of the LMC, following \mci\ and \mcii\ (see text, and Appendix
    \ref{sec:lmcanalysis}). C-rich AGB stars are shown in blue, O-rich
    in red, and others in black. The axes indicate linear distances
    (in kpc) in the disc plane of the LMC. The two circles are at 5
    and 10 kpc from the adopted LMC centre. Right top: empirical CDF
    of the C-rich and O-rich distributions of radial distances from
    the LMC centre. Shown as dotted lines are polynomials of fourth
    degree, fitted to the observed CDFs. Right, second from top:
    fractional residuals $|$CDF-fit$|$/CDF. Right, third from top: the
    derivatives $\partial$CDF/$\partial d$ of the fitted polynomials,
    i.e. the PDF of the distributions of radial distances from the LMC
    centre, up to 10~kpc. The PDFs are normalized to unit area. Also
    shown are normalized histograms of the radial distances, obtained
    using the \emph{Bayesian Blocks} method, which guarantees optimal
    binning. Right bottom: C:O ratio (green) as a function of radial
    distance $d$. It was obtained by dividing the two (unnormalized)
    PDFs from the panel above, and uses the left vertical axis. The
    C:M count ratio found by \citet[][see their fig.~6]{Feast+2010} is
    shown in orange, and uses the right vertical axis. Both vertical
    axes span the range from 0 to $1.1\ \times$ the maximum of the C:O
    or C:M curves, respectively. The curves are reliable only up to
    distances of $\sim$6~kpc, beyond which the uncertainties are very
    large due to small source number counts per radial annulus. See
    Section~\ref{sec:coratiolmc} for details.}
  \label{fig:co-ratio-lmc}
\end{figure}

To investigate the spatial variation of the C:O ratio in LMC, we
analyse it as a function of radial distance from the LMC centre. We
apply the coordinate transformations laid out in a two-paper series on
the structure of the Magellanic Clouds, \citet{vdMarel_Cioni_MC1_2001}
and \citet{vdMarel_MC2_2001} (hereafter \mci\ and \mcii). These
transformations de-project the disc of the LMC to a face-on view and
thus enable the measurement of the C:O ratio in concentric rings
around the LMC centre. See Appendix \ref{sec:lmcanalysis} for the
exact procedure.  Fig.~\ref{fig:co-ratio-lmc} shows in its left top
panel the stars in the LMC in equatorial coordinates. The samples
after de-projecting and applying our \WISE\ colour and magnitude
selection criteria for C-rich and O-rich AGB stars are shown in the
left bottom panel. C-rich stars are shown in blue ($n_{\rm C} =
1114$), O-rich in red ($n_{\rm O} = 238$), and other stars as black
dots ($n = 330$). The two grey rings are at 5 and 10~kpc radial
distance from the LMC centre, and in the LMC disc. The two
distributions of radial distances are indeed different. A two-sided
Kolmogorov--Smirnov test of the two samples yields a $p$-value
$\approx 6.5\times10^{-6}$, allowing us to reject the hypothesis that
both are drawn from the same underlying distribution at a highly
significant level. In fact, if we knew that both samples were
distributed \emph{normally}, with their respective means and
variances, this $p$-value would afford rejection with $\sqrt{2}
\erfinv (1-p) = 4.5\sigma$ confidence.

To test for the existence of a radial gradient in the C:O ratio, we
construct empirical cumulative distribution functions (CDF) of the
radial distances of the C-rich and O-rich stars from the LMC
centre. The CDFs are shown as blue and red lines in the right top
panel of Fig. \ref{fig:co-ratio-lmc}, and are defined as ${\rm CDF}(d)
= \sum_{i=1}^n\! 1\{d_i<d\}$. This function increases by one at each
observed distance $d_i$ in the sorted list of distances, i.e. it
counts the number of stars observed up to a distance $d$. A proper CDF
would need to be normalized by $1/n$, but we refrained from doing so
in the right top panel, to show how the absolute member counts in the
two samples differ. To obtain analytic representations of the CDFs we
fitted them with polynomials of various degrees, and also with the
CDFs of Normal distributions. The best fits could be obtained with
fourth degree polynomials, which were significantly better than third
degree polynomials and Gaussians. The fitted polynomial function is
$P_4 \equiv \sum_{i=0}^4 a_i x^i$, with coefficients $a_i = (-76.37,
256.03, 6.38, -4.54, 0.26)$ for the CDF of C-rich stars, and $a_i =
(-30.77, 107.91, -15.00, 0.65, 0.01)$ for O-rich stars (lowest order
first). The fits are plotted with dotted lines in the top right panel
of Fig.~\ref{fig:co-ratio-lmc}.

The second panel from the top on the right shows the fractional
residuals of our fits (their absolute values). They are generally
well-controlled at a level $<$5 per cent, except in the innermost
region. The third panel from the top shows as red and blue lines the
probability density functions (PDF) of the two samples, obtained as
derivatives $\partial {\rm CDF}/\partial d$ of the fitted CDFs and
normalized to unit area. Also shown are histograms of the
distributions of radial distances, which were obtained using Bayesian
Blocks. The striking similarity of the PDFs obtained by polynomial
fits of the empirical CDFs, and the histograms computed
\emph{independently} via Bayesian Blocks, demonstrates that the
results are not strongly affected by methodology. Encouraged by this
agreement, we plot in the right bottom panel of
Fig. \ref{fig:co-ratio-lmc} the ratio of the (unnormalised) PDFs of
the C-rich and O-rich AGB samples, i.e. the radial C:O number count
ratio in the LMC. The value increases from $\sim$3 in the innermost
regions to $\la$19 at about 6~kpc, before turning down sharply. This
downturn is probably not reliable because the number counts of both
C-rich and O-rich stars are quite small at these distances. In fact,
\citet{Feast+2010}, who re-analysed the \emph{DENIS} data of
\citet{Cioni2009} by removing unpopulated areas in the LMC sky, find a
similar trend in the C:O star count ratio (C:M star ratio, to be more
precise), and furthermore derive large uncertainties for the data
points beyond $\sim$7~kpc (see their fig.~6).

It is reassuring that both the \citet{Feast+2010} results and our
analysis show the same trend of increasing C:O ratio with distance
from the LMC centre. Nevertheless, there are significant
differences. Between 2 and 6~kpc from the centre, the region where the
results of both studies can be considered reliable,
\citeauthor{Feast+2010} find a variation from approximately 0.25 to
0.5 while we find much larger ratios that range from $\sim$4 to 18, a
larger dynamic range as well. As discussed above, the two
determinations of the C:O ratio involve different properties. The
\citeauthor{Cioni2009} sample used by \citeauthor{Feast+2010} was
selected from the photospheric emission of AGB stars while our
\WISE\-selected samples are based on the radiation from the dusty
shells that typify late stages of evolution of these stars. That a
much larger C:O ratio is determined from dust re-radiation than from
photospheric emission implies a higher efficiency of dust formation
for the C-rich stars. A detailed comparison of the two results or,
equivalently, predicting the variation of the C:O ratio for dusty AGB
stars from the variation observed for early AGB stars, would require
detailed modelling of AGB evolution (e.g.,
\citealt{Marigo2008,Girardi2010}). In addition to prescriptions for
AGB evolution of both O-rich and C-rich stars, their age and
metallicity distributions and the dust formation efficiency also need
to be known or assumed. While such an effort is beyond the scope of
this study, it has the potential to shed new light on the formation
and evolution of the LMC.

%
%

%
%

%
\begin{table*}
\begin{minipage}[b]{0.63\hsize}
  \begin{center}
    \caption{Number ratio of C-to-O AGB stars in the Galactic
      vicinity. The circular regions selected in
      Fig.~\ref{fig:co-ratio} and their properties are listed in lines
      1--10. Valid patches are those that contain both C-rich and
      O-rich stars.}\label{tab:coratio}
      \begin{tabular}{rlrrrrr}
        \hline
        No. & Quantity                        & Full sky     & GC    & Sgr DSph   & SMC    & LMC \\
        \hline
         1  & Galactic longitude (deg)        & n/a          & 0.    &       5.57 & 302.80 & 280.47     \\
         2  & Galactic latitude (deg)         & n/a          & 0.    &     -14.17 & -44.30 & -32.89     \\
         3  & Radius (deg)                    & n/a          & 8.06  &       5.62 &   4.00 &   8.06     \\
         4  & Number of C-rich stars n(C)     & 1629         & 60    &         21 &    161 &   1079     \\
         5  & Number of O-rich stars n(O)     & 14\,969      & 4500  &         62 &     24 &    235     \\
         6  & Number of valid \textsc{Healpix} patches & 196 & 21    &          7 &      3 &     14     \\[5pt]
            & \underline{$\log_{10}$ n(C)\,/\,n(O)} & & & & & \\
         7  & For entire selection            &        -0.96 & -1.88 &      -0.47 &   0.83 & \ph{-}0.66 \\
         8  & Mean of valid patches           &        -0.85 & -1.71 &      -0.19 &   0.84 & \ph{-}0.87 \\
         9  & Min of valid patches            &        -2.61 & -2.61 &      -0.90 &   0.48 &      -0.40 \\
        10  & Max of valid patches            &   \ph{-}1.38 & -0.70 & \ph{-}0.48 &   1.32 & \ph{-}1.38 \\
        \hline
      \end{tabular}
  \end{center}
\end{minipage}
\end{table*}
%

\section{Modelling \WISE\ colours}
\label{sec:models}

Why do objects segregate in colour space and what causes the
particular segregation patterns of different families of \WISE\
sources? These questions were already addressed by \cite{IE2000} in
the context of \IRAS\ sources. The dust radiative transfer problem
possesses general scaling properties, as discussed in detail by
\cite{IE1997}. There are only two input quantities whose magnitudes
matter: the dust optical depth at some fiducial wavelength and its
temperature at some position in the source. All other input is defined
by dimensionless, normalized profiles that describe (1) the spectral
shapes of the dust absorption and scattering coefficients, (2) the
spectral shape of the external heating radiation, and (3) the dust
spatial distribution. Physical dimensions, e.g. luminosity and linear
sizes, are irrelevant. Scaling applies to arbitrary geometries and
implies that {\em different dusty objects are expected to segregate in
  CC diagrams primarily because their dust shells are different, not
  because they have different central sources}. This behaviour is
displayed by the objects in the \IRAS\ PSC and is similarly expected
for \WISE\ colours.

We now describe our basic modelling framework for dusty stellar
environments and then discuss comparisons of models with data in more
detail. All modelling was performed with our public code \D\
\citep{DUSTY}.

\subsection{Model properties}

%
\begin{figure}
  \includegraphics[width=\hsize]{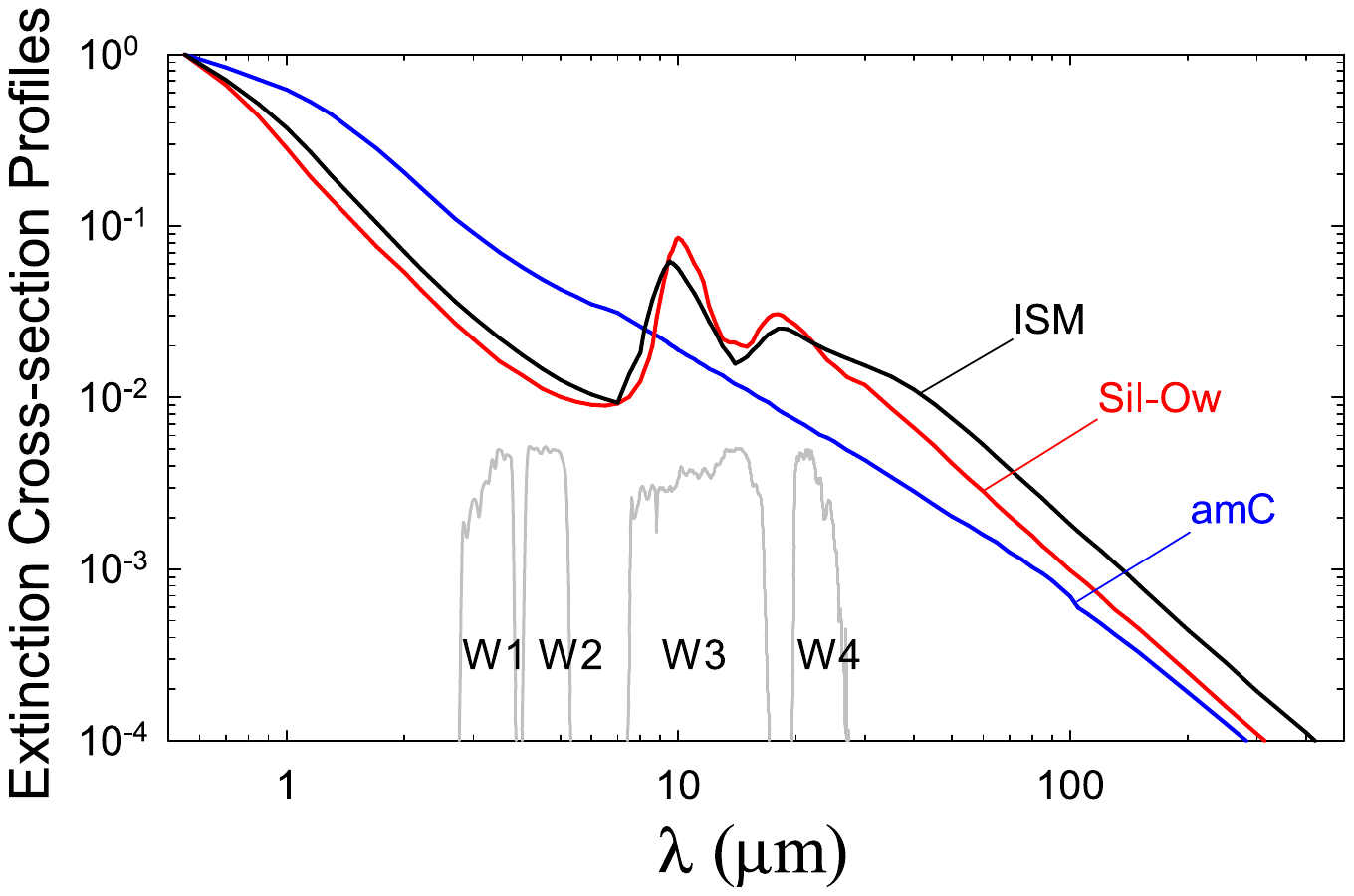}
  \caption{Extinction cross-sections, normalized to unity at
    0.55~\mic, of model dust for spherical grains with an MRN size
    distribution. Black: standard ISM dust mixture from \citet{Oss92}
    and \citet{Draine03}. Red: warm silicates from \citet{Oss92}
    (Oss-w). Blue: amorphous carbon from \citet{Hann88} (amC). The
    shapes of the \WISE\ filter response functions are outlined in
    light grey.}
  \label{fig:Xsections}
\end{figure}
%
We employ two families of dust grains. Our `standard ISM' dust
contains a standard Galactic mix of 53 per cent silicates, with
optical constants from the \cite{Oss92} `cold dust' for silicate, and
47 per cent graphite, with optical properties from
\cite{Draine03}. The other family represents pure-composition dust
typically found in AGB stars. Dust around C-rich stars comprises
amorphous carbon, whose properties we take from \cite{Hann88}, while
for O-rich stars we take the warm silicates from \cite{Oss92}. We
assumed spherical grains and considered both single-size dust grains
and models with MRN grain-size distribution \citep*{MRN}. \D\ handles
dust distributions through the composite grain approximation by
averaging over the chemical composition and size distribution. This
approach was shown to reproduce reasonably accurately the results from
exact treatment of grain mixtures \citep[see e.g.][]{EfRR94}. The
extinction profiles used in the calculations are shown in
Fig.~\ref{fig:Xsections}.

%
\begin{figure}
  \includegraphics[angle=0,width=1.\hsize]{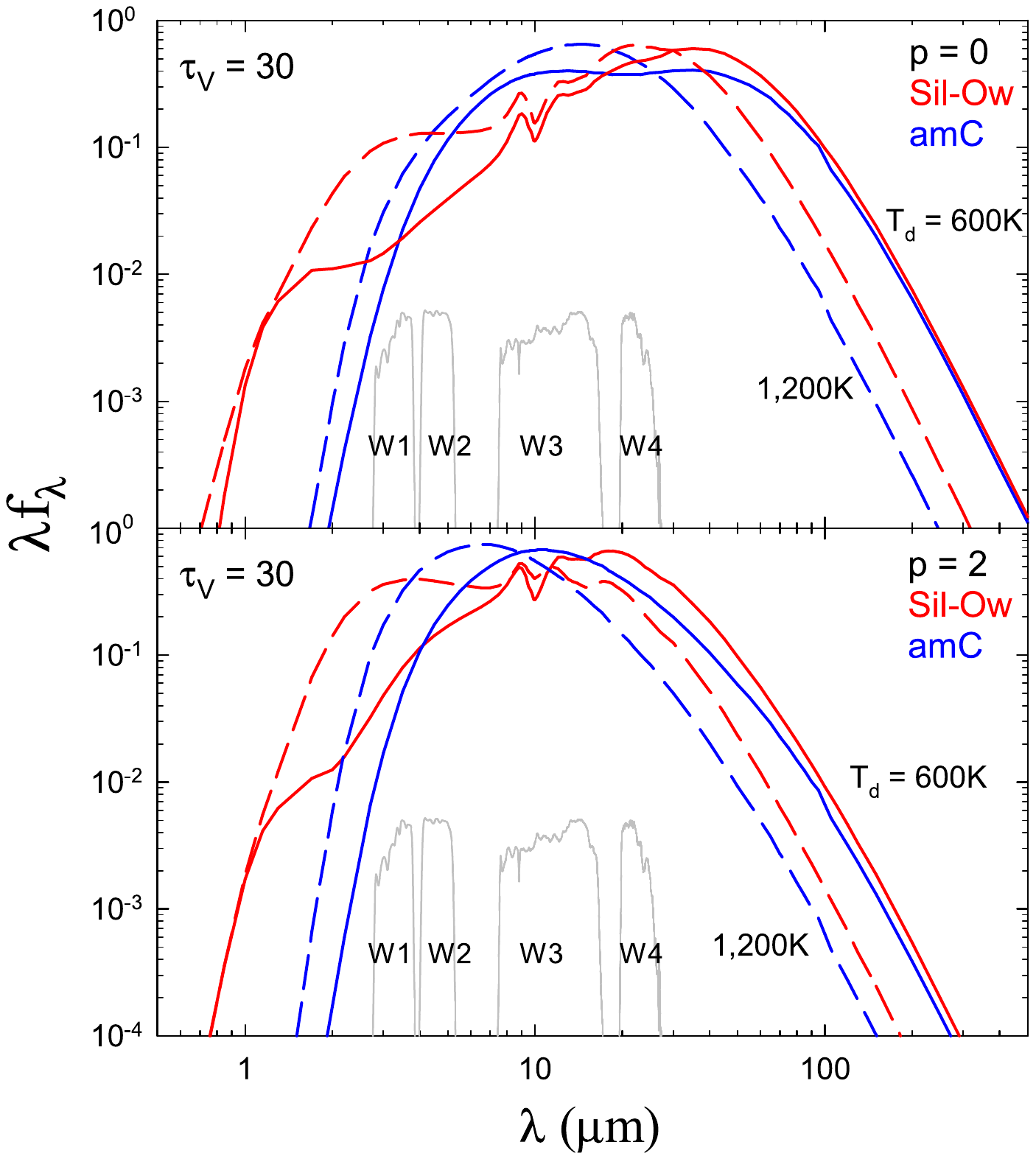}
  \caption{Model SEDs for spherical dust shells, centrally heated by a
    10\,000~K source. Silicate dust models are shown in red, amorphous
    carbon in blue. In both cases the grains follow the MRN size
    distribution and the extinction profiles are shown in
    Fig.~\ref{fig:Xsections}. The shell optical depth is \tV\ = 30 at
    visual, its relative thickness is $Y = 100$ and its temperature on
    the inner boundary is either $T_d=600$~K (solid lines) or
    $T_d=1200$~K (dashed). The density profiles are power-law
    $r^{-p}$ with $p$ = 0 and 2, as marked. The shapes of the \WISE\
    filter response functions are outlined in light grey.}
  \label{fig:spectra-amCSil}
\end{figure}
%

With \D\ we produced models of spherical dust shells around central
sources; the code calculates exact solutions for the dust temperature
profiles, emerging flux and other properties of interest. As noted
above, the only input parameters whose magnitudes matter are the dust
temperature on the shell inner boundary and the shell overall optical
depth at a given wavelength; all other model input describes
dimensionless, normalized profiles. We computed a range of dusty shell
models with overall optical depth at visual from \tV\ = 0.1 to 200 and
dust temperature on the inner boundary from $T_d$ = 300~K to 1200~K.
The spectral shape of the central source was taken as a blackbody with
temperature $T_s=2500$~K, 5000~K and 10\,000~K. For the radial density
distribution of the spherical dusty shell we used a power-law $r^{-p}$
with $p$ = 0, 1 and 2. In addition, for the AGB shells we used the
radial density profiles determined from a solution of the full
equation of motion of radiatively driven dusty winds, obtained with
the RDW option of \D\ \citep[see][]{EI01}. The shell relative
thickness, ratio of outer to inner radius $Y \equiv R_{\rm out}/R_{\rm
  in}$, was taken as 100. Resulting SEDs for some representative cases
are shown in Fig.~\ref{fig:spectra-amCSil}.

\subsection{Colour tracks}
\label{sec:tracks}

As noted above (Section~\ref{sec:naked_stars}), a naked star
corresponds to a point near the origin of the \WISE\ colour
space. Surrounding the star with a dust shell will shift the colours
to the red, moving the source away from the origin. Given the shell
properties and varying only its overall optical depth, the dust
temperature profile remains unchanged as long as the shell remains
optically thin. Therefore the colours remain the same; all optically
thin dusty shells with the same properties correspond to a single
point in the colour space.  Increasing the optical depth further, the
shell temperature profile starts to change when it enters the
optically thick regime, leading to deviation of the source colours
away from the optically-thin point towards redder colours. This
results in a track in the 3D \WISE\ colour space, with position along
the track determined by the shell overall optical depth
\tV. Projections on the 2D colour planes produce tracks in CC
diagrams, like those shown in the right panel of
Fig.~\ref{fig:compare-to-wise}. This figure shows that the tracks for
$p$ = 0 and 1 shells of ISM dust trace the boundaries of the region
occupied by extragalactic sources in the \s[1,2]--\s[2,3] CC
diagram. This behaviour is in agreement with the findings by
\cite{Levenson07} and \cite{Sirocky+2008} that the deep silicate
absorption features observed in ULIRG spectra require burial of
luminous central sources in very optically thick ($\tV \ga 100$) dusty
shells with flat ($p$ = 0--1) radial density distributions. The figure
also shows that stars occupy a region much closer to the colour origin
so that the optical thicknesses of their dust shells are generally
much smaller.

We now discuss in detail the \WISE\ colours of Galactic
sources. Tabulations of the colour tracks used in figures displayed
below are provided in Table \ref{tab:tabulations}, where we also list
the model in-band fluxes, useful for studies which depend on source
luminosity.

%
\begin{figure}
  \center
  \includegraphics[width=\hsize]{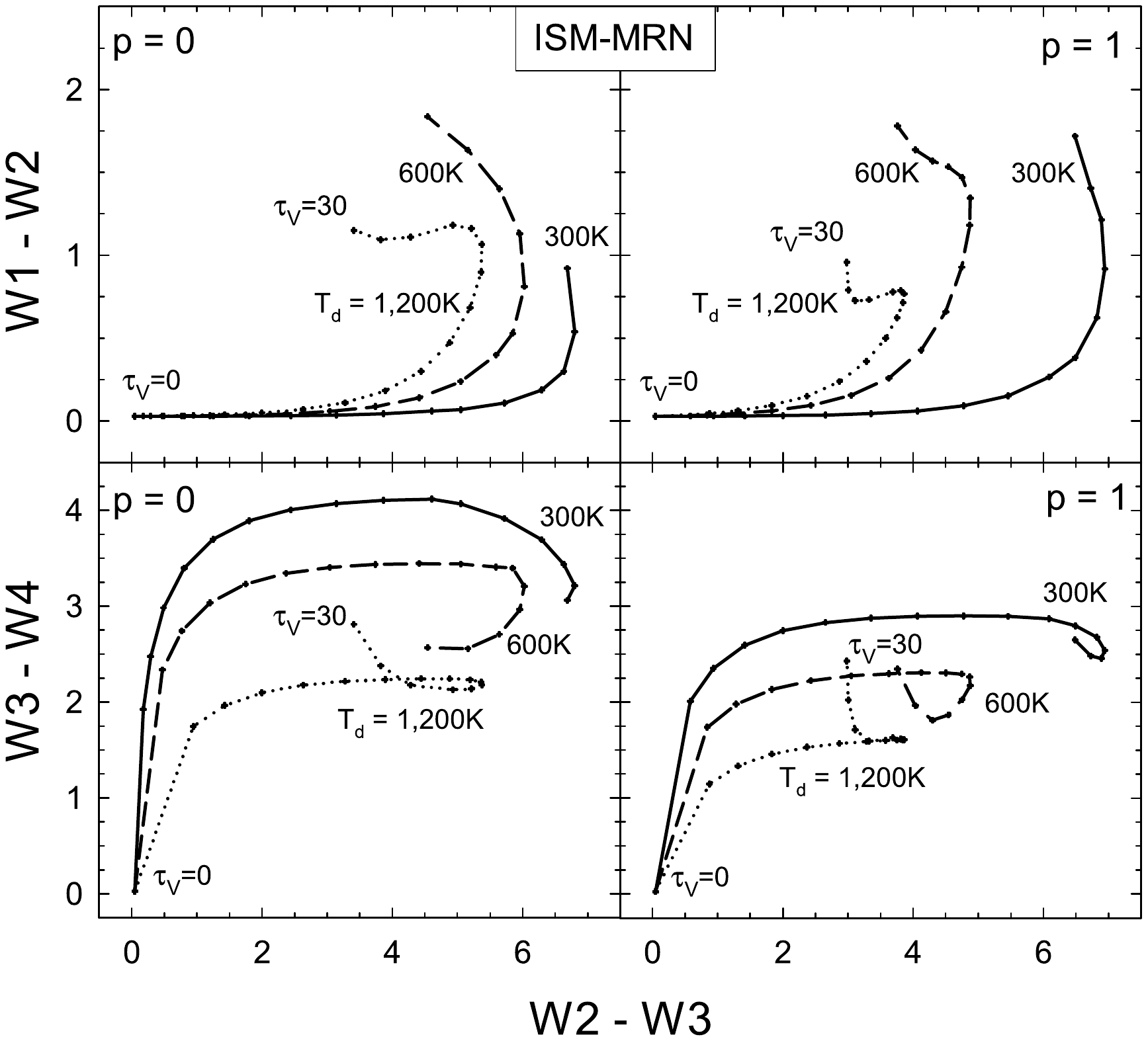}
  \caption{Model colour tracks for spherical dusty shells around a
    central blackbody source with $T_s=10\,000$~K. The shell density
    profile is a power-law $r^{-p}$ with $p=0, 1$, as marked. Each
    track starts at the colours of the dust-free blackbody and the
    optical depth increases along the track from \tV$=0$ to
    \tV$=30$. The dust is taken as ISM type with MRN size
    distribution.  In each panel different tracks correspond to dust
    temperature $T_d$ on the shell inner boundary of 300~K (solid
    lines), 600~K (dashed lines) and 1200~K (dotted lines).}
  \label{fig:D-colors}
\end{figure}
%

\subsection{YSO tracks}
\label{sec:yso}

\cite{IE2000} found that \IRAS\ colours of YSOs could be explained by
the colour tracks of dust shells with flat ($p = 0$) density
distributions. Fig.~\ref{fig:D-colors} explores in detail the \WISE\
colours expected from $p$ = 0 and 1 radial profiles for spherical
shells of ISM dust with MRN size distribution and optical depths up to
$\tV = 30$, heated by a central 10\,000~K blackbody. Models were
produced for dust temperatures on the shell inner boundary varying
from $T_d=300$~K to $T_d=1200$~K. This figure provides a zoom-in on
the smaller-\tV\ region of the tracks for shallow density
distributions that were shown in Fig.~\ref{fig:compare-to-wise}. Its
various plots show the range of possibilities for colour tracks
covering the likely region of parameter space for dust shells around
YSOs. A characteristic feature of flat density distributions is that
even a small increase in optical depth can significantly enhance long
wavelength emission because it places more dust at large radial
distances with low temperatures. For this reason, the $\tV < 1$
portions of the tracks are parallel to the colour axis with longer
wavelengths while maintaining a value of roughly 0 for the other axis
with the `warmer' colour. Deviations from this behaviour increase with
$T_d$ and are mostly noticeable in the \s[2,3]--\s[3,4] diagram,
especially for the $p$ = 1 tracks.
%
\begin{figure}
  \includegraphics[angle=0,width=1.\hsize]{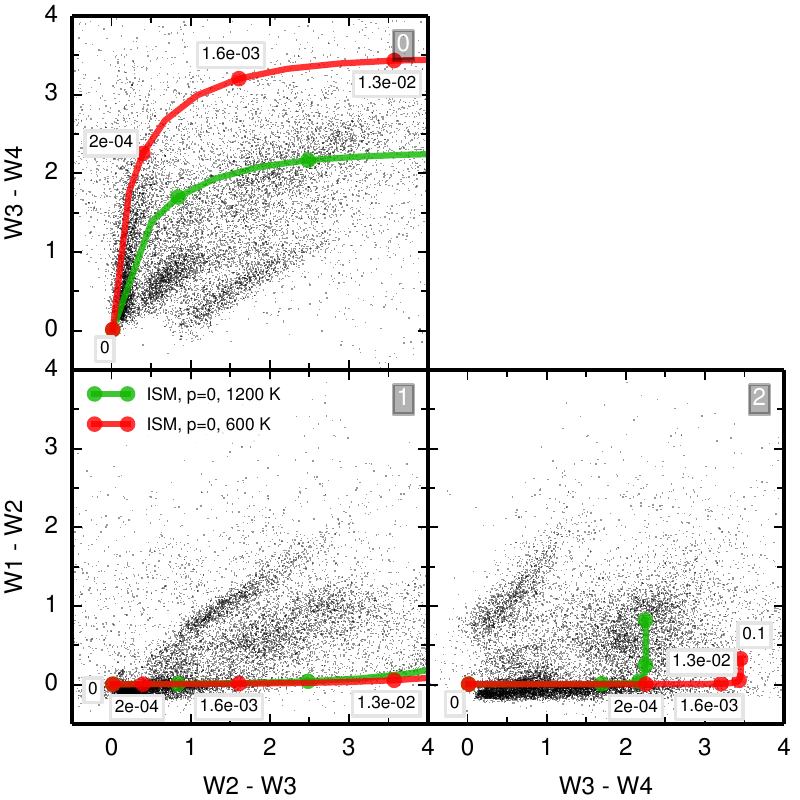}
  \caption{Colour distributions of Galactic sources, as in
    Fig.~\ref{fig:star-plumes}; each panel is numbered at its top
    right corner. Overplotted are model colour tracks of spherical
    dusty shells with standard ISM dust grains and uniform density
    profiles ($p$ = 0). The shells surround central stars with
    Vega-like emission. The dust temperature on the shell inner
    boundary is either 600~K (red tracks) or 1200~K (green). Position
    along the track depends only on the dust optical depth; a few
    values of the $V$-band optical depth are marked. The tracks
    delineate the plume of YSOs.}
\label{fig:plume_yso}
\end{figure}
%

Fig.~\ref{fig:plume_yso} shows the comparison with observations. Its
three panels reproduce the projections on to 2D colour planes of the
3D colour distribution of bright \WISE\ sources that were shown
previously in Fig.~\ref{fig:star-plumes}, with model tracks plotted
over the source distribution. The models producing these tracks are
the same as the $p$ = 0 models shown in Fig.~\ref{fig:D-colors} except
for the spectral shape of the heating source: instead of a pure
blackbody they use the empirical Vega fit employed for normalization
by \citet{Wright+2010} (see also Section~\ref{sec:naked_stars}). These
tracks follow rather well the long, narrow plume of stars along the
\s[2,3] and \s[3,4] axes at \s[1,2] $\approx 0$ in Panels~1 and 2 of
Fig.~\ref{fig:plume_yso} and along the \s[3,4] axis at \s[2,3]
$\approx 0$ in Panel~0. This plume has been previously identified as
the locus of YSOs among bright \WISE\ sources
(Section~\ref{sec:envelopes}). In most colour-colour combinations,
little differentiates the tracks for different values of $T_d$, the
dust temperature on the inner boundary. The tracks diverge
significantly only in the panels involving the \s[3,4] colour
combination, where only the $T_d$ = 1200~K track passes through the
blobs roughly clustered around (2.5,2) in Panel~0 and (2.2,0.7) in
Panel~2. It is not clear, though, that these blobs give a reliable
indication of thermal dust emission because these sources could be
dominated by PAH emission covered by the broad \w[3] filter (see
Section~\ref{sec:agb}). A reliable determination of the nature of
these sources would require in-depth, detailed modelling of individual
objects.

The small range of optical depths spanned by the main plume of SF
sources implies that \WISE\ has uncovered a large number of YSO with
small amounts of dust around them while giving little indication of
such objects covered by deep layers of dust. This is surprising
because heavy dust obscuration is expected in many YSOs. Indeed, the
SF regions identified by \IRAS\ were characterized by large optical
depths (\tV\,\about\,1--10; see fig.~3 in \citealt{IE2000}). These
differences reflect the starkly different sensitivity limits of the
two facilities. \IRAS\ required for detection strong IR emission,
which precluded sources with optically thin dusty shells. However,
optically thick YSO shells, which were a prerequisite for \IRAS\
detection, produce \WISE\ colour differences that exceed 4. Such large
differentials are eliminated by the cuts we place on saturation
limits, which remove sources too bright at short wavelengths, and on
signal-to-noise ratio, which remove sources too weak at the long
wavelength bands. As is evident from Fig.~\ref{fig:compare-to-wise},
the \s[2,3] colour of optically thick $p$ = 0 objects is larger than
4, entering the domain populated by extragalactic sources which we
deliberately sought to avoid in this analysis.  Relaxing our selection
criteria should therefore reveal a heavily obscured YSO population
mixed with the extragalactic sources.

\begin{figure}
  \includegraphics[angle=0,width=1.\hsize]{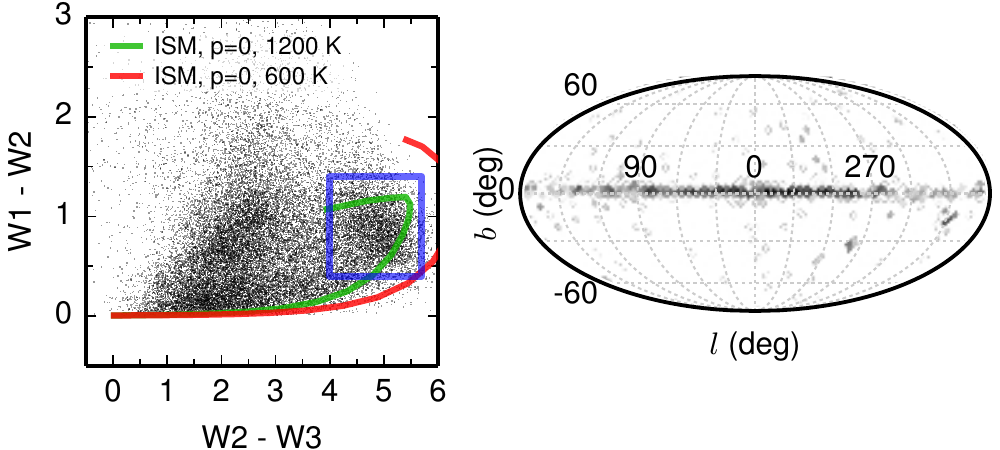}
  \caption{Identifying YSOs in the \WISE\ data base. Left:
    distribution of all sources with $10 \le \w[1] \le 11$ in a \WISE\
    CC diagram. All other cuts are as in Fig.~\ref{fig:plume_yso},
    except that there is no constraint on Galactic latitude. The green
    and red \D\ model colour tracks are as in
    Fig.~\ref{fig:plume_yso}. The blue rectangle outlines the expected
    colour region of YSOs with high dust obscuration. Right: all-sky
    distribution of the sources in the blue rectangle (in Mollweide
    projection), with the grey-scale normalized to peak source count
    density. It is heavily concentrated towards the Galactic plane,
    where local SF regions are predominantly found.}
  \label{fig:yso_extended}
\end{figure}

To test this hypothesis we identified in our basic selections those
sources that are moderately bright in \w[1] and are matched by high
optical depths of our model colour tracks. These would most likely be
YSOs with high shell optical depths. Such a population, compatible
with the previous \IRAS\ detections, indeed exists in the \WISE\ data
base. It can be found approximately within $10 \lesssim \w[1] \lesssim
11$, $0.4 \lesssim \s[1,2] \lesssim 1.4$, and $4 \lesssim \s[2,3]
\lesssim 5.7$. All sources with \mbox{$10 \le \w[1] \le 11$} are shown
in the CC scatter diagram in Fig.~\ref{fig:yso_extended} (left panel);
all other selection criteria from Fig.  \ref{fig:plume_yso} apply,
except that the constraint on Galactic latitude was dropped. The CC
diagram contains 29\,892 sources while the blob outlined with a blue
rectangle contains 6192 objects, presumably YSOs with high dust
optical depths. Indeed, the centre of the selection cube, $\w[1] =
10.5$, $\s[1,2] = 0.9$, $\s[2,3] = 4.85$, implies that $\w[3] = 4.75$
(Vega). With this magnitude we can use $F_\nu = F_\nu^{\rm iso} \times
10^{{\rm -m_{Vega}/2.5}}$ to compute the in-band flux density $F_\nu$,
where $F_\nu^{\rm iso}$ is the zero-magnitude flux density in
Jansky. It is given as 31.674~Jy for band 3 in \citet[][see their
table 1]{Jarrett+2011}. Thus our \w[3] magnitude ${\rm m_{Vega}} =
4.75$ corresponds to $\approx 0.4$~Jy at 12~\mic\ (\w[3]) while the
faint limit of the \IRAS\ Faint Source Catalogue at 12~\mic\ is
0.2~Jy. Therefore, our findings are compatible with \IRAS\ discovering
only sources in the blue box but not those with lower 12~\mic\
fluxes. Also plotted in the left panel of Fig.~\ref{fig:yso_extended}
are the same model colour tracks from Panel~1 of
Fig.~\ref{fig:plume_yso}, but over an expanded colour range. The green
track ($T_d = 1200$~K) shows a turn-over at $\s[2,3] \sim 5.5$, and
all its \tV\ values between 0.025 and 9.4 stay perfectly within the
blue YSO rectangle. The right panel of Fig. \ref{fig:yso_extended}
shows the all-sky distribution of the $\sim 6200$ sources from the YSO
plume. The distribution is extremely heavily concentrated towards the
Galactic plane, a preferred location to find local SF regions. This is
further evidence that our conclusions are consistent. The choice of
dusty shell models for YSOs can be debated because some young stars,
e.g. T Tauri stars, are known to be surrounded by dusty
disks. \citet{Vinkovic+2003} showed that a flared disk produces an SED
identical to that of an equivalent optically thin spherical dust
shell. The latter is thus a good proxy for the more realistic disk
model. There is no way to distinguish between them, or any combination
of the two, with flux measurements alone; imaging is necessary for
that. From equations~7 and 8 in section 2.1 of
\citeauthor{Vinkovic+2003}, a constant density shell (as applied in
our YSO models) is equivalent to a disk with linear flaring. The very
small optical depths we find in the WISE data of YSOs may be also
interpreted as the difference in flaring angle across the disk
(Equation 7), but it is difficult to decide without imaging.

%
%

\subsection{AGB tracks and model parameter constraints}
\label{sec:agb-analysis}

\begin{figure}
  \includegraphics[angle=0,width=1.\hsize]{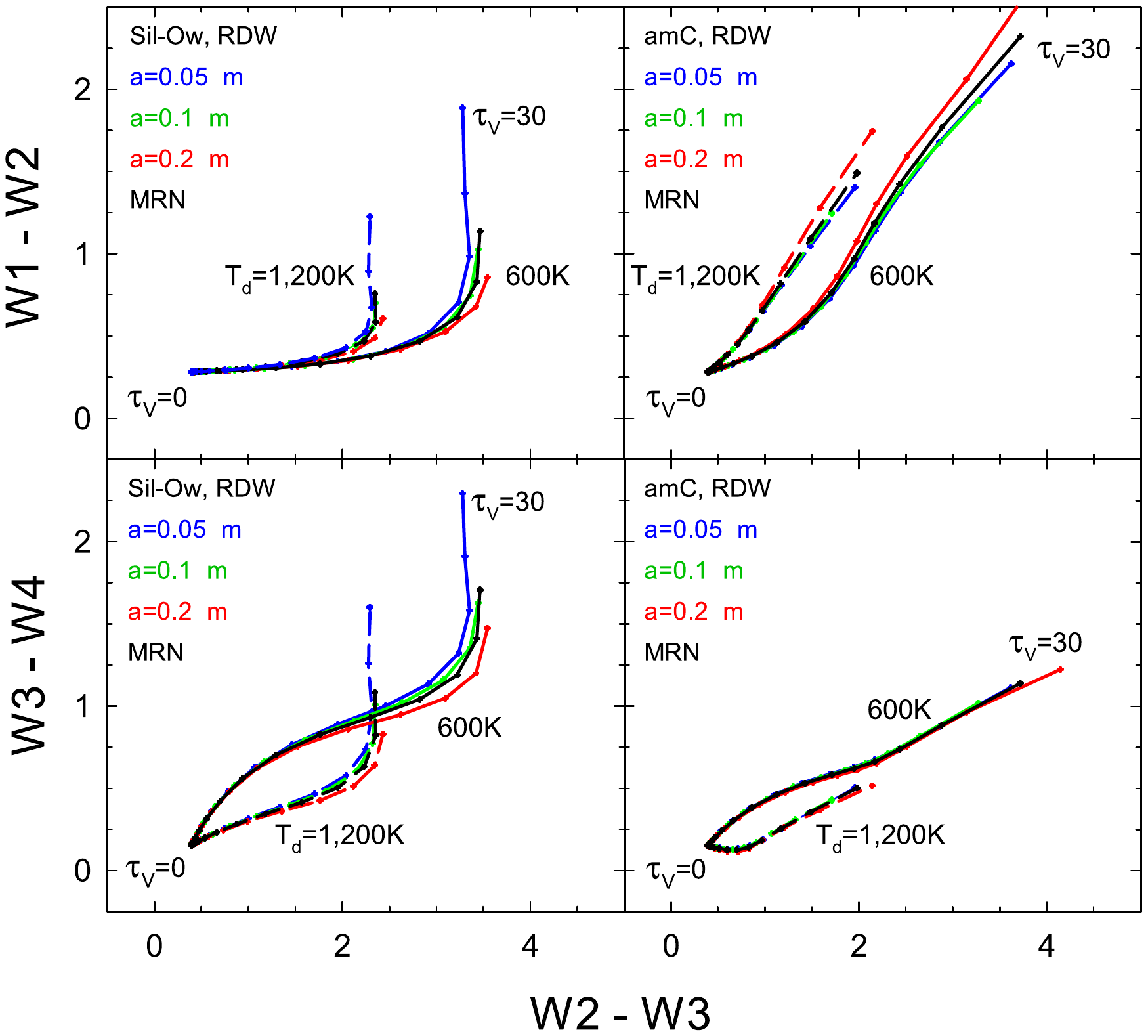}
  \caption{Model colour tracks for dusty RDW around a central star
    with $T_s$ = 2500~K. Left panels are for silicate dust (Sil-Ow),
    right panels -- for amorphous carbon (amC). The dust temperature
    $T_d$ at the wind origin is taken as 600~K (solid lines) and
    1200~K (dashed lines). Model tracks for various grain sizes and
    for the MRN size distribution are colour coded as marked. The
    overall optical depth of the shell increases along each track from
    \tV\ = 0 to \tV\ = 30.}
  \label{fig:RDW}
\end{figure}

Radiatively driven winds (RDW) have been shown to well describe
infrared emission from dusty envelopes around AGB stars \citep[see,
e.g.:][and references therein]{EI01, IE10}.  Their circumstellar dust
distribution approximately follows a power-law $1/r^2$.  \D\ can
calculate RDW models exactly and in Fig.~\ref{fig:RDW} we show RDW
model colour tracks for a central source at $T_s=2500$~K and a $Y=100$
shell. Tracks are shown for two temperatures at the inner dust
boundary, $T_d=600$~K and $T_d=1200$~K.  AGB stars are classified as
O-rich or C-rich stars (see Section~\ref{sec:agb}) and models were
produced with two corresponding types of dust: warm silicates from
\citet{Oss92} (Sil-Ow, left panels), and amorphous carbon (amC, right
panels). Both single-size grains and an MRN size distribution were
used, as indicated in the legend. As seen from Fig.~\ref{fig:RDW}, the
effect of different grain sizes on the colour tracks is negligible
compared to varying the dust temperature.

%
%

Amorphous carbon absorbs more in the \w[1] and \w[2] bands, while
silicates absorb more in the \w[3] and \w[4] bands (see
Fig.~\ref{fig:Xsections}). Therefore, given two circumstellar shells
with the same overall optical depth but different chemical
composition, the silicate shell is expected to be redder in the
\c[2,3] colour and bluer in the \c[1,2] colour. As illustrated in
Fig.~\ref{fig:RDW}, the \WISE\ colour diagrams can separate sources
with steep density profile and predominantly silicate dust
(e.g. typical AGB stars) from those with carbonaceous dust
(e.g. carbon stars), for the same dust sublimation temperature.

How do the RDW model tracks compare to AGB colours measured by \WISE?
First, we can fix the stellar temperature $T_s$ at 2500~K and the
ratio of inner-to-outer radius of the shells at 100, since both are
quite typical values for AGB stars. We also investigated other $T_S$,
but they did not agree as well with the data (the allowed temperature
range is only a few hundred K wide). The two most important free model
parameters are dust interface temperature, $T_d$, and shell optical
depth, \tV. This two-dimensional space of model parameters maps to
\WISE\ colour space (and vice versa).
%
\begin{figure}
  \includegraphics[width=\hsize]{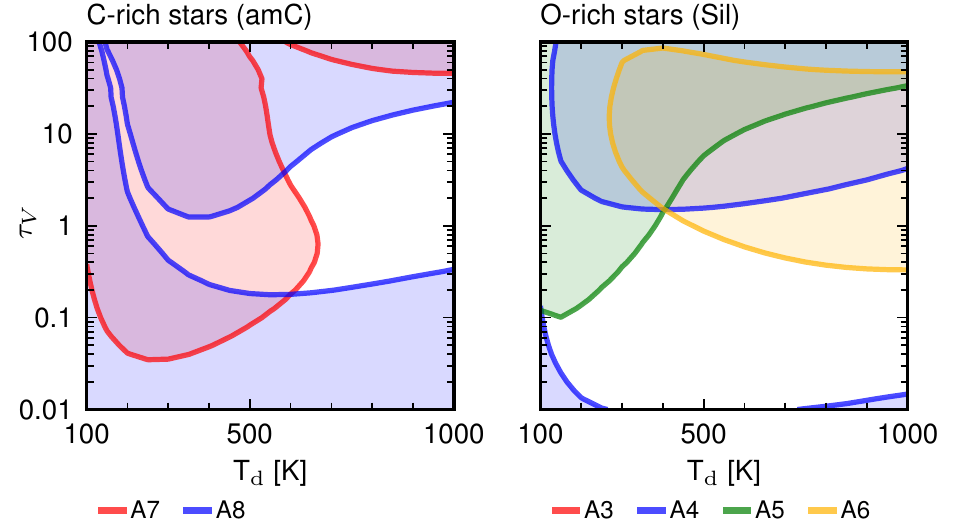}
  \caption{Translation of our AGB colour selection criteria [defined
    in equations~\eqref{eq:O1}-\eqref{eq:C2}] into model parameter
    constraints. Left for C-rich AGB stars, right for O-rich
    stars. Each equation translates into curved areas in the
    ($T_d$,\tV) parameter space. The excluded areas are shaded (see
    legend for the colour/equation code). Only the white fraction of
    each panel contains permissible model parameter combinations. See
    text for details.}
  \label{fig:agb-analysis-exclusion}
\end{figure}
%

We have demonstrated in Section~\ref{sec:agb} that AGB stars can be
selected using only \WISE\
photometry. Equations~\eqref{eq:O1}--\eqref{eq:C2} in Appendix
\ref{app:agb-selection} are the mathematical definitions of our
selection criteria. The corresponding loci of AGB stars are visualized
in Fig.~\ref{fig:agb-selection} with red and blue lines. These
equations also define lower and upper bounds of several \WISE\ linear
colour combinations. For instance, equation \eqref{eq:C1} can be
written as \mbox{$x - 0.629\, y \in [-0.198,0.359]$} (where \mbox{$x$
  = \c[1,2]}, \mbox{$y$ = \c[2,3]}). With given dust properties (for
example, amorphous carbon dust), we can compute the value of this
linear colour combination, $x - 0.629\, y$, for any \D\ model
parameter combination ($T_d$,\tV). Then the colour selection criterion
demands that this value be within $[-0.198,0.359]$. Therefore, the
interval boundaries are simply contours in the parameter plane, and
are plotted as solid lines in Fig.~\ref{fig:agb-analysis-exclusion}
(for amorphous carbon dust in the left panel, for O-rich star models
in the right panel). The legends relate the equation numbers to
colours used for plotting. The areas outside of the interval are
excluded by the selection criterion, and are shown shaded in the
figure. Each of the colour-selection equations
\mbox{\eqref{eq:O1}--\eqref{eq:C2}} translates to such
parameter-exclusion areas. The complement of their union is the only
portion of the parameter space with permissible combinations. It is
shown in white in Fig.~\ref{fig:agb-analysis-exclusion}.

Since our C-rich star colour selection volume is confined quite
tightly (cf. Fig. \ref{fig:agb-selection}), the allowed parameter
space for \D\ models is also well-constrained. The dust shells must
have interface temperatures not lower than $\simeq 600$~K and the
visual optical depths are in the range $\approx 0.2-20$. For O-rich
star models the selection in CC space was not as tight; consequently
the permissible parameter ranges are less well-constrained. Our
results seem to indicate that C-rich AGB stars tend to have higher
maximal optical depths than O-rich AGBs, and that they lack the very
lowest \tV\ values. In contrast, O-rich AGB models extend to smaller
optical depths, and values larger than a few are excluded. This
conclusion is supported by samples analysed in \citet{IE10}; see their
figs~5 and 6, with data from \citet{Young1995},
\citet{Richards_Yates1998} and \citet{Olofsson+1993} (we caution the
reader that those data sets may be affected by unknown selection
effects). We emphasize that our results should be viewed qualitatively
and note that the parameter boundaries are `soft' because they depend
on the exact colour selection criteria.

\begin{figure*}
  \includegraphics[width=0.7\hsize]{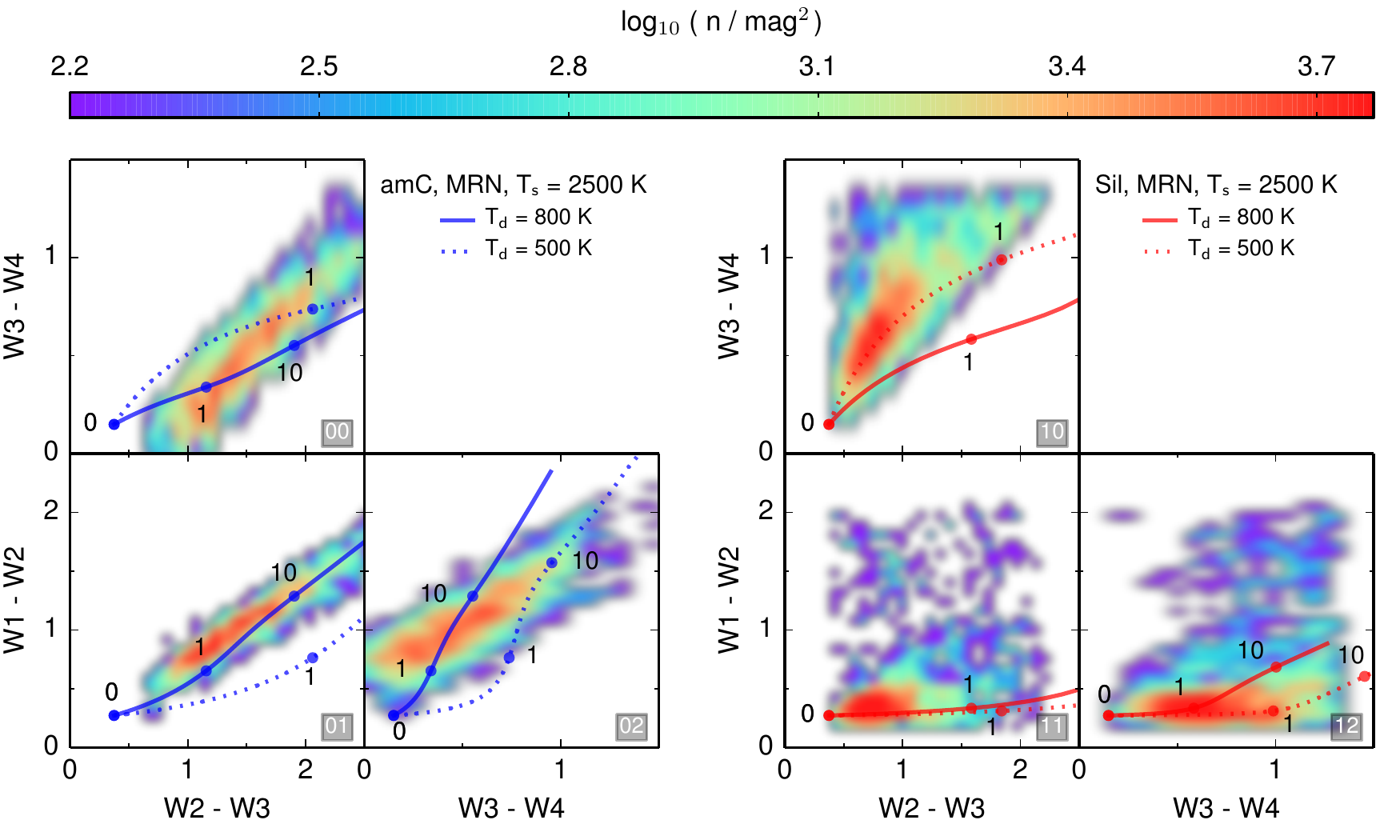}
  \caption{Colours of AGB stars measured by \WISE, and fiducial \D\
    model colour tracks for RDW. The number distribution of observed
    AGB colours is shown with a logarithmic colour scale for C-rich
    (left panels) and O-rich AGB stars (right panels). Overplotted as
    solid and dotted lines are fiducial \D\ model colour tracks
    computed for both chemistries, with two dust interface
    temperatures that bracket the observed colours (see legend). The
    total radial optical depth of the shell models is printed at some
    positions along the tracks.}
  \label{fig:agb-analysis-cc}
\end{figure*}

The distribution of C-rich and O-rich AGB stars in the \WISE\ colour
space, as previously selected in Section~\ref{sec:agb} and
Appendix~\ref{app:agb-selection}, is shown in
Fig.~\ref{fig:agb-analysis-cc}.  Overplotted are fiducial RDW model
colour tracks, for which the temperature at the inner boundary is
$T_d=800$~K (solid lines) or $T_d=500$~K (dotted), in accordance with
the parameter ranges found above. All tracks are computed for a shell
thickness of 100, and with a stellar temperature of $T_S=2500$~K, as
explained before. Panels~00--02 (left side) show our C-rich star
selection and colour tracks computed with amorphous carbon dust, while
Panels~10--12 (right) show our O-rich star selection and tracks with
`warm' circumstellar silicates from \citet{Oss92}.  The model tracks
do not match all AGB plumes in all three orthogonal projections
simultaneously, but the range spanned by models of these dust
interface temperatures brackets well the observed \WISE\ colours of
AGB stars.

%
%

The $T_d=800$~K RDW model with amorphous carbon dust (blue solid line)
follows the plume of C-rich stars reasonably well in all three
two-dimensional projections of the \WISE\ colour space
(Panels~00--02). Although the agreement is not perfect, it is likely
that these 0.2--0.3~mag large offsets could be rectified with minor
modifications of the dust opacity. We have not attempted a detailed
study because such small changes could also be explained by deviations
from spherical symmetry, non-uniform dust distribution, variations in
photospheric temperature as the star ascends the asymptotic giant
branch sequence, and possibly other effects. These degeneracies cannot
be broken with only three \WISE\ colours. The red tracks in
Panels~10--12 of Fig.~\ref{fig:agb-analysis-cc} correspond to a
fiducial model for O-rich stars with standard silicate dust opacity
and $T_d$~=~800~K and 500~K. Models in this temperature range bracket
the 3D distribution of O-rich AGBs in the \WISE\ colour volume. It is
remarkable that most of the extension of this O-rich star plume can be
covered with shell optical depths below a value of unity.

\section{Summary and discussion}
\label{sec:summary}

With detections for about 560 million objects, \WISE\ represents a
major step forward in the surveying and understanding of the infrared
sky. Its combination of up to 1000 times higher sensitivity than
previous IR surveys, and a relatively high angular resolution, enables
studies of source samples with unprecedented statistical
significance. We have computed colour tracks in the \WISE\ photometric
system for two families of models: smooth spherical shells and clumpy
dust torus models, and decided to study local/Galactic populations in
this paper, and the extragalactic sky in Nikutta et al. (in
preparation).

The model tracks provide guidance for the observed distribution of
sources in \WISE\ CC diagrams, such as illustrated in the left panel
in Fig.~\ref{fig:compare-to-wise}. The main results of our
computations are shown in the right panel in
Fig.~\ref{fig:compare-to-wise}.  The model-based colour tracks outline
reasonably well the distribution of sources expected to be associated
with dust emission (i.e. all source types except T dwarfs and stars
without dusty shell). The location of dust-free stars coincides very
well with the optically thin point of all dust-shell models
irrespective of the dust composition. Deviations from the zero-point
are caused by \WISE\ filters \w[1] and \w[2] being sensitive to the
blackbody peaks of the coolest stars, thus producing colours other
than those for the Rayleigh--Jeans slope.  In the other extreme, at
high optical depths, the \D\ model tracks of ISM dust shells with
shallow radial density distributions overlap with those CC areas
populated by Ultra-Luminous Infrared Galaxies (ULIRGs), which are
known to be deeply shrouded in dust \citep{Sirocky+2008}.

We find that \WISE\ photometry alone is sufficient to reliably
classify C-rich and O-rich asymptotic giant branch stars with dusty
shells (Section~\ref{sec:agb}). With the large population of AGB stars
in the \WISE\ Catalogue we were able to define magnitude and colour
selection criteria that distinguish between the two classes. It is
necessary to regard the distribution of \WISE\ colours as a
three-dimensional arrangement in colour--colour--colour space. The
traditional two-dimensional orthogonal projections are very helpful,
but not sufficient. With the ability to cleanly separate the local
populations of C-rich and O-rich AGB stars we computed the most
detailed map of the C:O-rich star count ratio for the Milky Way and
its satellites (see Section~\ref{sec:co-ratio}). We find a strong
spatial gradient away from the Galactic plane, with the Magellanic
Clouds showing a C:O ratio about 1000 times higher than the Galactic
bulge. We also clearly uncover the Sagittarius Dwarf Spheroidal Galaxy
in the map. We find that the C:O dusty AGB star ratio increases with
distance from the LMC centre about twice as fast as measured for
near-IR selected samples of early AGB stars. Detailed modelling of
these measurements has the potential to shed new light on the
formation and evolution of LMC.

\D\ model tracks for shells with RDWs follow the identified AGB plumes
well. By mapping the previously found colour selection criteria to
model parameters, specifically inner dust shell temperature $T_d$ and
shell optical depth \tV, we found observational constraints on these
model parameters. They seem to indicate slight differences in the
physical properties of C-rich and O-rich AGB shells (see
Section~\ref{sec:agb-analysis}).

Spherical dusty shell models with flat radial density distributions
generate \WISE\ colours compatible with YSOs. YSO samples selected
using \IRAS\ and \WISE\ surveys have very little overlap due to vastly
different imaging depths: while \IRAS\ was sensitive to bright sources
with relatively high visual optical depths, \WISE\ detects much
fainter YSOs with relatively small optical depths (see
Section~\ref{sec:yso}).

To investigate further how well models describe the \WISE\ colours of
IR sources, sufficiently large samples need to be studied
object-by-object, with SED fitting for each source over a large
wavelength range. The \Spitzer\ SAGE Survey \citep{Meixner+2006}, for
instance, has been previously used for such a task.
\citet{Riebel+2012} matched the IR \Spitzer\ spectra with optical and
near-IR data to produce broad-band SEDs for more than 30\,000 AGB
stars in the LMC, and fitted them with \emph{GRAMS} models
\citep{Sargent+2011_GRAMS, Srinivasan+2011_GRAMS}. Thanks to \WISE, a
similar study can be now undertaken for AGB stars in the Galaxy.

\C\ models of AGN dust torus emission, with their more complicated
geometry and dust distributions, are much better than simple spherical
shells in explaining the general location of QSOs and other AGNs in
the \WISE\ CC diagram. The location of the highest \C\ model density
aligns almost perfectly with the centre of the circular region
identified as `Seyferts' by \citeauthor{Wright+2010} in the left panel
of Fig. \ref{fig:compare-to-wise}. A detailed study of the \WISE\
colours of AGNs will be presented in the companion paper
(\citetalias{WISE2}).


\paragraph*{Acknowledgements:}

We are grateful to Maria-Rosa Cioni for illuminating discussions and
for making her data available to us. We reproduced fig.~12 from
\citet{Wright+2010} as part of our Fig.~\ref{fig:compare-to-wise} by
permission of the AAS and the authors. We thank Chao-Wei Tsai for
providing us a high-quality version. We thank both referees for their
helpful suggestions which improved the paper. We express our gratitude
to Xavier Koenig for reminding us of the disk/shell dichotomy of YSO
models (see Section \ref{sec:yso}), and to James Davenport for his
insightful post-submission comments on the variability information in
the WISE database. We are also thankful to Patricia Ar\'{e}valo and
Ron Wilhelm for helpful discussions. RN acknowledges support by the
ALMA-CONICYT fund, project No. 31110001, and by FONDECYT grant
No. 3140436. \v{Z}I thanks the Hungarian Academy of Sciences for its
support through the Distinguished Guest Professor grant
No. E-1109/6/2012. \v{Z}I and NHW acknowledge support by NSF grant
AST-1008784 to the University of Washington. ME acknowledges NASA
support and the award of an NPP Senior Fellowship from ORAU, which
supported a sabbatical leave at JPL/Caltech where much of this work
was completed. This publication makes use of data products from the
Wide-field Infrared Survey Explorer, which is a joint project of the
University of California, Los Angeles, and the Jet Propulsion
Laboratory/California Institute of Technology, funded by the National
Aeronautics and Space Administration. Funding for SDSS-III has been
provided by the Alfred P. Sloan Foundation, the Participating
Institutions, the National Science Foundation, and the U.S. Department
of Energy Office of Science. The SDSS-III web site is
\url{http://www.sdss3.org/}. SDSS-III is managed by the Astrophysical
Research Consortium for the Participating Institutions of the SDSS-III
Collaboration including the University of Arizona, the Brazilian
Participation Group, Brookhaven National Laboratory, Carnegie Mellon
University, University of Florida, the French Participation Group, the
German Participation Group, Harvard University, the Instituto de
Astrofisica de Canarias, the Michigan State/Notre Dame/JINA
Participation Group, Johns Hopkins University, Lawrence Berkeley
National Laboratory, Max Planck Institute for Astrophysics, Max Planck
Institute for Extraterrestrial Physics, New Mexico State University,
New York University, Ohio State University, Pennsylvania State
University, University of Portsmouth, Princeton University, the
Spanish Participation Group, University of Tokyo, University of Utah,
Vanderbilt University, University of Virginia, University of
Washington, and Yale University. This research has made use of the NED
which is operated by the Jet Propulsion Laboratory, California
Institute of Technology, under contract with the National Aeronautics
and Space Administration. Some of the results in this paper have been
derived using the \textsc{Healpix}/\textsc{Healpy} package
\citep{HEALPIX} .



\begin{thebibliography}{}

\bibitem[\protect\citeauthoryear{{Baldwin}, {Phillips}, \&
  {Terlevich}}{{Baldwin} et~al.}{1981}]{BPT}
{Baldwin}, J.~A., {Phillips}, M.~M.,  \& {Terlevich}, R. 1981, \pasp, 93, 5

\bibitem[\protect\citeauthoryear{{Battinelli} \& {Demers}}{{Battinelli} \&
  {Demers}}{2011}]{BD2011}
{Battinelli}, P.,  \& {Demers}, S. 2011, in ASP Conf. Ser., Vol. 445, Why
  Galaxies Care about AGB Stars II: Shining Examples and Common Inhabitants,
  ed. F.~{Kerschbaum}, T.~{Lebzelter}, \& R.~F. {Wing}, 479

\bibitem[\protect\citeauthoryear{{Blanton} et~al.}{{Blanton}
  et~al.}{2005}]{vagc}
{Blanton}, M.~R., et~al. 2005, \aj, 129, 2562

\bibitem[\protect\citeauthoryear{{Bolatto} et~al.}{{Bolatto}
  et~al.}{2007}]{Bolatto2007}
{Bolatto}, A.~D., et~al. 2007, \apj, 655, 212

\bibitem[\protect\citeauthoryear{{Boyer} et~al.}{{Boyer}
  et~al.}{2013}]{Boyer+2013}
{Boyer}, M.~L., et~al. 2013, \apj, 774, 83

\bibitem[\protect\citeauthoryear{{Cioni}}{{Cioni}}{2009}]{Cioni2009}
{Cioni}, M.-R.~L. 2009, \aap, 506, 1137

\bibitem[\protect\citeauthoryear{{Cioni} \& {Habing}}{{Cioni} \&
  {Habing}}{2003}]{CH2003}
{Cioni}, M.-R.~L.,  \& {Habing}, H.~J. 2003, \aap, (CH03), 402, 133

\bibitem[\protect\citeauthoryear{{Covey} et~al.}{{Covey} et~al.}{2007}]{Covey}
{Covey}, K.~R., et~al. 2007, \aj, 134, 2398

\bibitem[\protect\citeauthoryear{{Dalcanton} et~al.}{{Dalcanton}
  et~al.}{2012}]{JD2012}
{Dalcanton}, J.~J., et~al. 2012, \apjs, 198, 6

\bibitem[\protect\citeauthoryear{{Draine}}{{Draine}}{2003}]{Draine03}
{Draine}, B.~T. 2003, \apj, 598, 1017

\bibitem[\protect\citeauthoryear{{Efstathiou} \& {Rowan-Robinson}}{{Efstathiou}
  \& {Rowan-Robinson}}{1994}]{EfRR94}
{Efstathiou}, A.,  \& {Rowan-Robinson}, M. 1994, \mnras, 266, 212

\bibitem[\protect\citeauthoryear{{Egan}, {Van Dyk}, \& {Price}}{{Egan}
  et~al.}{2001}]{Egan2001}
{Egan}, M.~P., {Van Dyk}, S.~D.,  \& {Price}, S.~D. 2001, \aj, 122, 1844

\bibitem[\protect\citeauthoryear{{Elitzur} \& {Ivezi{\'c}}}{{Elitzur} \&
  {Ivezi{\'c}}}{2001}]{EI01}
{Elitzur}, M.,  \& {Ivezi{\'c}}, {\v Z}. 2001, \mnras, 327, 403

\bibitem[\protect\citeauthoryear{{Feast}, {Abedigamba}, \& {Whitelock}}{{Feast}
  et~al.}{2010}]{Feast+2010}
{Feast}, M.~W., {Abedigamba}, O.~P.,  \& {Whitelock}, P.~A. 2010, \mnras, 408,
  L76

\bibitem[\protect\citeauthoryear{{Girardi} et~al.}{{Girardi}
  et~al.}{2010}]{Girardi2010}
{Girardi}, L., et~al. 2010, \apj, 724, 1030

\bibitem[\protect\citeauthoryear{{G{\'o}rski} et~al.}{{G{\'o}rski}
  et~al.}{2005}]{HEALPIX}
{G{\'o}rski}, K.~M., {Hivon}, E., {Banday}, A.~J., {Wandelt}, B.~D., {Hansen},
  F.~K., {Reinecke}, M.,  \& {Bartelmann}, M. 2005, \apj, 622, 759

\bibitem[\protect\citeauthoryear{{Hanner}}{{Hanner}}{1988}]{Hann88}
{Hanner}, M. 1988, {Grain optical properties}, Technical report

\bibitem[\protect\citeauthoryear{{Ita} et~al.}{{Ita} et~al.}{2008}]{Ita2008}
{Ita}, Y., et~al. 2008, \pasj, 60, 435

\bibitem[\protect\citeauthoryear{{Ivezi{\'c}}, {Beers}, \&
  {Juri{\'c}}}{{Ivezi{\'c}} et~al.}{2012}]{IBJ2012}
{Ivezi{\'c}}, {\v Z}., {Beers}, T.~C.,  \& {Juri{\'c}}, M. 2012, \araa, 50, 251

\bibitem[\protect\citeauthoryear{{Ivezi{\'c}} et~al.}{{Ivezi{\'c}}
  et~al.}{2014}]{astroMLText}
{Ivezi{\'c}}, {\v Z}., {Connolly}, A., {Vanderplas}, J.,  \& {Gray}, A. 2014,
  Statistics, Data Mining and Machine Learning in Astronomy (Princeton
  University Press)

\bibitem[\protect\citeauthoryear{{Ivezi\'{c}} \& {Elitzur}}{{Ivezi\'{c}} \&
  {Elitzur}}{1997}]{IE1997}
{Ivezi\'{c}}, {\v Z}.,  \& {Elitzur}, M. 1997, \mnras, 287, 799

\bibitem[\protect\citeauthoryear{{Ivezi{\'c}} \& {Elitzur}}{{Ivezi{\'c}} \&
  {Elitzur}}{2000}]{IE2000}
{Ivezi{\'c}}, {\v Z}.,  \& {Elitzur}, M. 2000, \apjl, 534, L93

\bibitem[\protect\citeauthoryear{{Ivezi{\'c}} \& {Elitzur}}{{Ivezi{\'c}} \&
  {Elitzur}}{2010}]{IE10}
{Ivezi{\'c}}, {\v Z}.,  \& {Elitzur}, M. 2010, \mnras, 404, 1415

\bibitem[\protect\citeauthoryear{{Ivezi{\'c}}, {Nenkova}, \&
  {Elitzur}}{{Ivezi{\'c}} et~al.}{1999}]{DUSTY}
{Ivezi{\'c}}, {\v Z}., {Nenkova}, M.,  \& {Elitzur}, M. 1999,
  arXiv:astro-ph/9910475

\bibitem[\protect\citeauthoryear{{Ivezi{\'c}} et~al.}{{Ivezi{\'c}}
  et~al.}{2007}]{Ivezic+2007}
{Ivezi{\'c}}, {\v Z}., et~al. 2007, \aj, 134, 973

\bibitem[\protect\citeauthoryear{{Jackson}, {Ivezi{\'c}}, \& {Knapp}}{{Jackson}
  et~al.}{2002}]{JIK2002}
{Jackson}, T., {Ivezi{\'c}}, {\v Z}.,  \& {Knapp}, G.~R. 2002, \mnras, 337, 749

\bibitem[\protect\citeauthoryear{{Jarrett} et~al.}{{Jarrett}
  et~al.}{2011}]{Jarrett+2011}
{Jarrett}, T.~H., et~al. 2011, \apj, 735, 112

\bibitem[\protect\citeauthoryear{{Javadi} et~al.}{{Javadi}
  et~al.}{2013}]{Javadi+2013}
{Javadi}, A., {van Loon}, J.~T., {Khosroshahi}, H.,  \& {Mirtorabi}, M.~T.
  2013, \mnras, 432, 2824

\bibitem[\protect\citeauthoryear{{Kazin} et~al.}{{Kazin}
  et~al.}{2010}]{Kazin2010}
{Kazin}, E.~A., et~al. 2010, \apj, 710, 1444

\bibitem[\protect\citeauthoryear{{Levenson} et~al.}{{Levenson}
  et~al.}{2007}]{Levenson07}
{Levenson}, N.~A., {Sirocky}, M.~M., {Hao}, L., {Spoon}, H.~W.~W., {Marshall},
  J.~A., {Elitzur}, M.,  \& {Houck}, J.~R. 2007, \apjl, 654, L45

\bibitem[\protect\citeauthoryear{{Marigo} et~al.}{{Marigo}
  et~al.}{2008}]{Marigo2008}
{Marigo}, P., {Girardi}, L., {Bressan}, A., {Groenewegen}, M.~A.~T., {Silva},
  L.,  \& {Granato}, G.~L. 2008, \aap, 482, 883

\bibitem[\protect\citeauthoryear{{Mathis}, {Rumpl}, \& {Nordsieck}}{{Mathis}
  et~al.}{1977}]{MRN}
{Mathis}, J.~S., {Rumpl}, W.,  \& {Nordsieck}, K.~H. 1977, \apj, 217, 425

\bibitem[\protect\citeauthoryear{{Meixner} et~al.}{{Meixner}
  et~al.}{2006}]{Meixner+2006}
{Meixner}, M., et~al. 2006, \aj, 132, 2268

\bibitem[\protect\citeauthoryear{{Menzies} et~al.}{{Menzies}
  et~al.}{2008}]{Menzies+2008}
{Menzies}, J., {Feast}, M., {Whitelock}, P., {Olivier}, E., {Matsunaga}, N.,
  \& {da Costa}, G. 2008, \mnras, 385, 1045

\bibitem[\protect\citeauthoryear{{Nenkova} et~al.}{{Nenkova}
  et~al.}{2008a}]{Nenkova+2008a}
{Nenkova}, M., {Sirocky}, M.~M., {Ivezi{\'c}}, {\v Z}.,  \& {Elitzur}, M.
  2008a, ApJ, 685, 147

\bibitem[\protect\citeauthoryear{{Nenkova} et~al.}{{Nenkova}
  et~al.}{2008b}]{Nenkova+2008b}
{Nenkova}, M., {Sirocky}, M.~M., {Nikutta}, R., {Ivezi{\'c}}, {\v Z}.,  \&
  {Elitzur}, M. 2008b, \apj, 685, 160

\bibitem[\protect\citeauthoryear{{Neugebauer} et~al.}{{Neugebauer}
  et~al.}{1984}]{IRAS1}
{Neugebauer}, G., et~al. 1984, \apjl, 278, L1

\bibitem[\protect\citeauthoryear{{Nikolaev} \& {Weinberg}}{{Nikolaev} \&
  {Weinberg}}{2000}]{NW2000}
{Nikolaev}, S.,  \& {Weinberg}, M.~D. 2000, \apj, 542, 804

\bibitem[\protect\citeauthoryear{{Nikutta} et~al.}{{Nikutta}
  et~al.}{2013}]{WISE2}
{Nikutta}, R., {Hunt-Walker}, N., {Nenkova}, M., {Ivezi{\'c}}, {\v Z}.,  \&
  {Elitzur}, M. 2013, \mnras, to be submitted (paper II)

\bibitem[\protect\citeauthoryear{{Obri{\'c}} et~al.}{{Obri{\'c}}
  et~al.}{2006}]{Obric}
{Obri{\'c}}, M., et~al. 2006, \mnras, 370, 1677

\bibitem[\protect\citeauthoryear{{Olnon} et~al.}{{Olnon} et~al.}{1986}]{IRAS2}
{Olnon}, F.~M., et~al. 1986, \aaps, 65, 607

\bibitem[\protect\citeauthoryear{{Olofsson} et~al.}{{Olofsson}
  et~al.}{1993}]{Olofsson+1993}
{Olofsson}, H., {Eriksson}, K., {Gustafsson}, B.,  \& {Carlstroem}, U. 1993,
  \apjs, 87, 305

\bibitem[\protect\citeauthoryear{{Ossenkopf}, {Henning}, \&
  {Mathis}}{{Ossenkopf} et~al.}{1992}]{Oss92}
{Ossenkopf}, V., {Henning}, T.,  \& {Mathis}, J.~S. 1992, \aap, 261, 567

\bibitem[\protect\citeauthoryear{{Peeters}, {Spoon}, \& {Tielens}}{{Peeters}
  et~al.}{2004}]{Peeters+2004}
{Peeters}, E., {Spoon}, H.~W.~W.,  \& {Tielens}, A.~G.~G.~M. 2004, \apj, 613,
  986

\bibitem[\protect\citeauthoryear{{Richards} \& {Yates}}{{Richards} \&
  {Yates}}{1998}]{Richards_Yates1998}
{Richards}, A.~M.~S.,  \& {Yates}, J.~A. 1998, Irish Astronomical Journal, 25,
  7

\bibitem[\protect\citeauthoryear{{Richards} et~al.}{{Richards}
  et~al.}{2003}]{Richards2003}
{Richards}, G.~T., et~al. 2003, \aj, 126, 1131

\bibitem[\protect\citeauthoryear{{Riebel} et~al.}{{Riebel}
  et~al.}{2012}]{Riebel+2012}
{Riebel}, D., {Srinivasan}, S., {Sargent}, B.,  \& {Meixner}, M. 2012, \apj,
  753, 71

\bibitem[\protect\citeauthoryear{{Sargent}, {Srinivasan}, \&
  {Meixner}}{{Sargent} et~al.}{2011}]{Sargent+2011_GRAMS}
{Sargent}, B.~A., {Srinivasan}, S.,  \& {Meixner}, M. 2011, \apj, 728, 93

\bibitem[\protect\citeauthoryear{{Scargle} et~al.}{{Scargle}
  et~al.}{2013}]{BayesianBlocks2013}
{Scargle}, J.~D., {Norris}, J.~P., {Jackson}, B.,  \& {Chiang}, J. 2013, \apj,
  764, 167

\bibitem[\protect\citeauthoryear{{Schneider} et~al.}{{Schneider}
  et~al.}{2010}]{Schneider2010}
{Schneider}, D.~P., et~al. 2010, \aj, 139, 2360

\bibitem[\protect\citeauthoryear{{Sirocky} et~al.}{{Sirocky}
  et~al.}{2008}]{Sirocky+2008}
{Sirocky}, M.~M., {Levenson}, N.~A., {Elitzur}, M., {Spoon}, H.~W.~W.,  \&
  {Armus}, L. 2008, \apj, 678, 729

\bibitem[\protect\citeauthoryear{{Smol{\v c}i{\'c}} et~al.}{{Smol{\v c}i{\'c}}
  et~al.}{2004}]{Smolcic+2004}
{Smol{\v c}i{\'c}}, V., et~al. 2004, \apjl, 615, L141

\bibitem[\protect\citeauthoryear{{Soifer}, {Neugebauer}, \& {Houck}}{{Soifer}
  et~al.}{1987}]{Soifer1987}
{Soifer}, B.~T., {Neugebauer}, G.,  \& {Houck}, J.~R. 1987, \araa, 25, 187

\bibitem[\protect\citeauthoryear{{Srinivasan}, {Sargent}, \&
  {Meixner}}{{Srinivasan} et~al.}{2011}]{Srinivasan+2011_GRAMS}
{Srinivasan}, S., {Sargent}, B.~A.,  \& {Meixner}, M. 2011, \aap, 532, A54

\bibitem[\protect\citeauthoryear{{Trams} et~al.}{{Trams}
  et~al.}{1999}]{Trams1999}
{Trams}, N.~R., et~al. 1999, \aap, 346, 843

\bibitem[\protect\citeauthoryear{{Tu} \& {Wang}}{{Tu} \&
  {Wang}}{2013}]{TuWang2013}
{Tu}, X.,  \& {Wang}, Z.-X. 2013, Research in Astronomy and Astrophysics, 13,
  323

\bibitem[\protect\citeauthoryear{{van der Marel}}{{van der
  Marel}}{2001}]{vdMarel_MC2_2001}
{van der Marel}, R.~P. 2001, \aj, (MC2), 122, 1827

\bibitem[\protect\citeauthoryear{{van der Marel} \& {Cioni}}{{van der Marel} \&
  {Cioni}}{2001}]{vdMarel_Cioni_MC1_2001}
{van der Marel}, R.~P.,  \& {Cioni}, M.-R.~L. 2001, \aj, (MC1), 122, 1807

\bibitem[\protect\citeauthoryear{{van der Veen} \& {Habing}}{{van der Veen} \&
  {Habing}}{1988}]{vdVeen}
{van der Veen}, W.~E.~C.~J.,  \& {Habing}, H.~J. 1988, \aap, 194, 125

\bibitem[\protect\citeauthoryear{{Vanderplas} et~al.}{{Vanderplas}
  et~al.}{2012}]{astroML}
{Vanderplas}, J., {Connolly}, A., {Ivezi{\'c}}, {\v Z}.,  \& {Gray}, A. 2012,
  in Conference on Intelligent Data Understanding (CIDU), 47

\bibitem[\protect\citeauthoryear{{Vijh} et~al.}{{Vijh} et~al.}{2009}]{Vijh2009}
{Vijh}, U.~P., et~al. 2009, \aj, 137, 3139

\bibitem[\protect\citeauthoryear{{Vinkovi{\'c}} et~al.}{{Vinkovi{\'c}}
  et~al.}{2003}]{Vinkovic+2003}
{Vinkovi{\'c}}, D., {Ivezi{\'c}}, {\v Z}., {Miroshnichenko}, A.~S.,  \&
  {Elitzur}, M. 2003, \mnras, 346, 1151

\bibitem[\protect\citeauthoryear{{Whitelock} et~al.}{{Whitelock}
  et~al.}{2009}]{Whitelock+2009}
{Whitelock}, P.~A., {Menzies}, J.~W., {Feast}, M.~W., {Matsunaga}, N.,
  {Tanab{\'e}}, T.,  \& {Ita}, Y. 2009, \mnras, 394, 795

\bibitem[\protect\citeauthoryear{{Wright} et~al.}{{Wright}
  et~al.}{2010}]{Wright+2010}
{Wright}, E.~L., et~al. 2010, \aj, 140, 1868

\bibitem[\protect\citeauthoryear{{York} et~al.}{{York} et~al.}{2000}]{SDSS}
{York}, D.~G.,  et~al. 2000, \aj, 120, 1579

\bibitem[\protect\citeauthoryear{{Young}}{{Young}}{1995}]{Young1995}
{Young}, K. 1995, \apj, 445, 872

\bibitem[\protect\citeauthoryear{{Zijlstra} et~al.}{{Zijlstra}
  et~al.}{2006}]{Zijlstra2006}
{Zijlstra}, A.~A., et~al. 2006, \mnras, 370, 1961

\end{thebibliography}


\appendix

\section{Selection criteria for AGB stars}
\label{app:agb-selection}

\begin{figure*}
  \center
  \includegraphics[width=1.5\columnwidth]{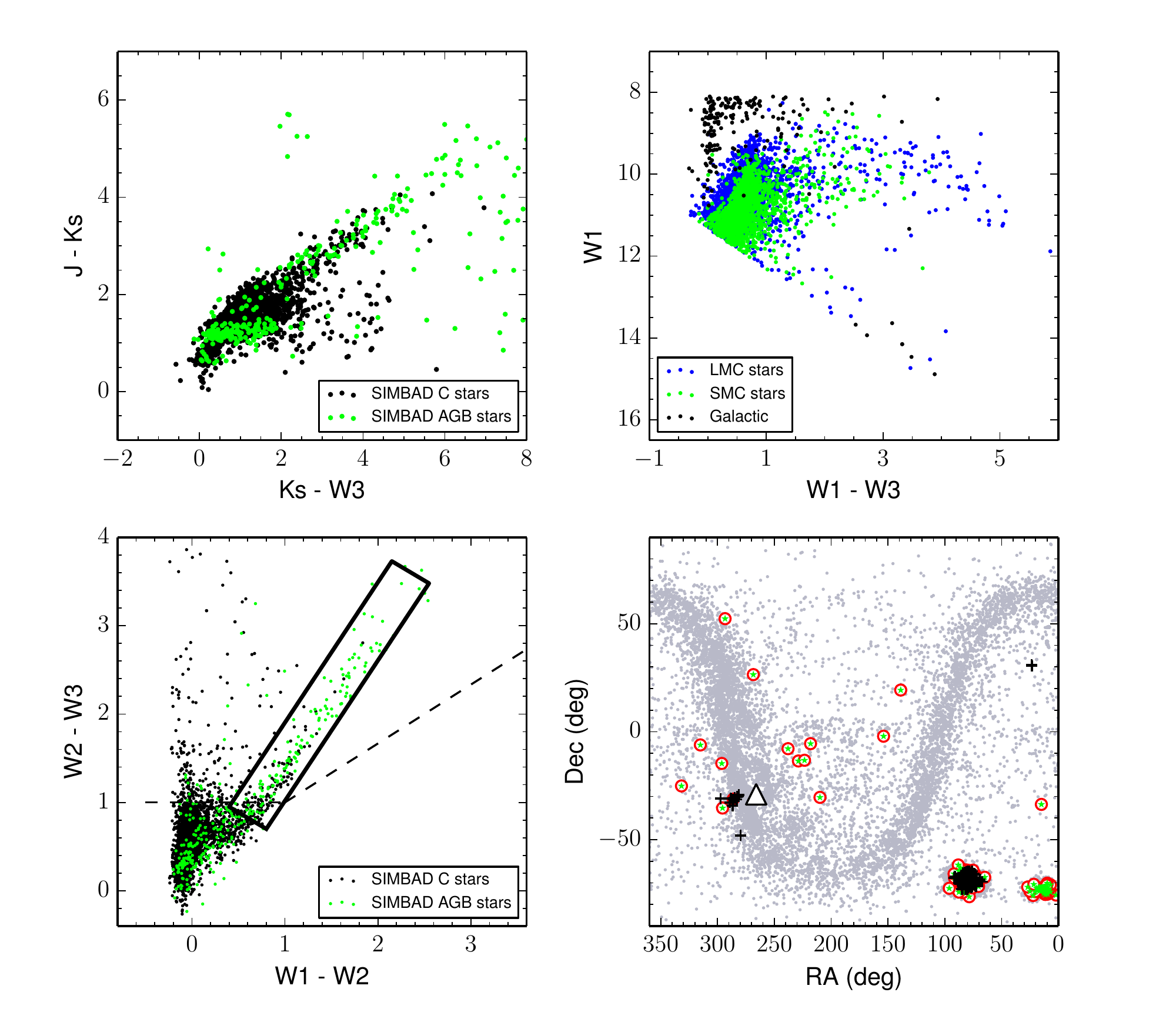}
  \caption{Distribution of \WISE\ sources whose classification listed
    in the SIMBAD data base is consistent with them being AGB
    stars. Top left panel: the $J-K_s$ versus $K_s-\w[3]$ diagram of
    \WISE\ sources identified in SIMBAD as C stars (black) and AGB
    stars (green, includes both O-rich and C-rich stars); similar to
    fig.~1 from \citet{TuWang2013}, except that saturated stars are
    not included, resulting in essentially all OH/IR stars being
    rejected. Top right panel: the \w[1] versus \c[1,3] diagram for
    all AGB stars in the Galaxy (black), LMC (blue), and SMC (green);
    similar to fig.~3 from \citet{TuWang2013}. Bottom left panel: the
    \c[2,3] versus \c[1,2] diagram for C stars (black) and AGB stars
    (green); similar to fig.~4 from \citet{TuWang2013}. The black box
    outlines a concentration of C-rich AGB stars. The dashed line was
    proposed by \citeauthor{TuWang2013} as a separator between OH/IR
    and C-rich AGB stars, but see text for our disagreement. Bottom
    right panel: the sky distribution in equatorial coordinates for
    stars within the black box in the bottom left panel, using the
    same colour scheme (green dots are encircled for better
    visibility). The entire sample of AGB, C, S, OH/IR, and Miras from
    SIMBAD with \WISE\ detection, but without magnitude limits are
    plotted in the background for reference; the distribution outlines
    the Galactic disc. The Galactic Centre is marked by the large
    triangle.}
  \label{fig:SIMBAD}
\end{figure*}
%

\begin{figure*}
  \center
  \includegraphics[width=1.5\columnwidth]{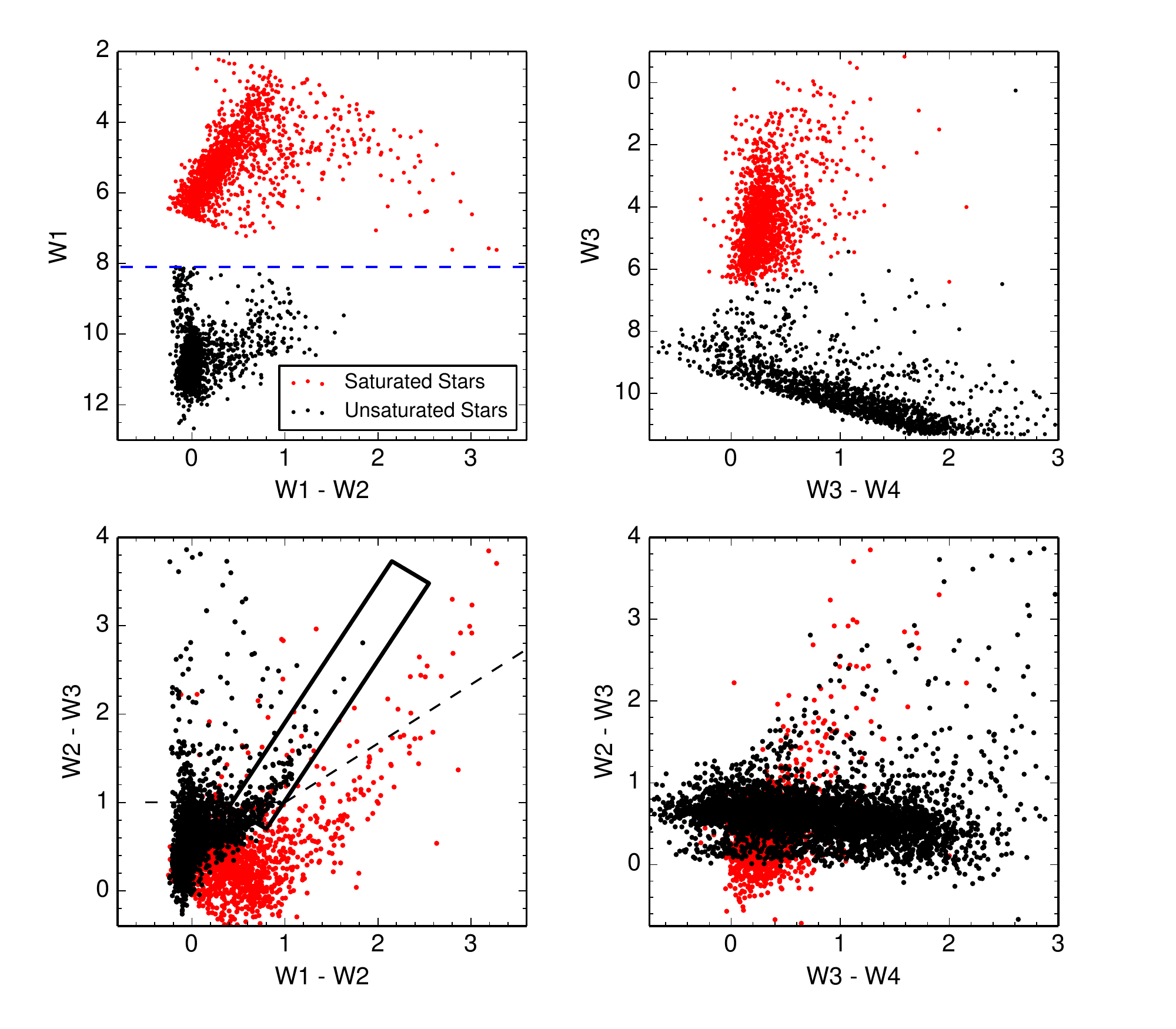}
  \caption{Similar to Fig. \ref{fig:SIMBAD}, but including saturated
    sources (with \w[1]$<$8.1 and \w[2]$<$6.7), which shown as red
    dots. Unsaturated stars are shown in black. In the upper left
    panel the blue dashed line shows the \w[1] saturation limit. In
    the lower left panel, the dashed line devised by
    \citet{TuWang2013} does not separate C-rich from O-rich AGB stars,
    but rather saturated from unsaturated ones.}
  \label{fig:saturated}
\end{figure*}
%

In order to develop selection criteria for AGB stars, we follow
\cite{TuWang2013} and analyse the \WISE\ colour distribution of known
AGB stars listed in the SIMBAD data base. Aided by their results, our
procedure is simpler. They first selected about 76 million sources
from the full \WISE\ All-Sky source catalogue with detections in
\w[1], \w[2] and \w[3] bands (and free of contamination and confusion
flags in all four bands) that have a unique 2MASS source within 3
arcsec. Then they positionally matched this sample to objects listed
in the SIMBAD data base within 2 arcsec (except in the case of high
proper-motion stars which were matched within 4 arcsec), and further
restricted the sample to 20 degrees from the Galactic plane, yielding
a sample of 3.2 million sources. This sample includes AGB stars, as
well as all other sources known to SIMBAD and detected by \WISE, which
satisfy the selection criteria.

We begin by obtaining a sample of stars from SIMBAD whose main
identifier was consistent with AGB stars (AGB star, C star, S star,
OH/IR star, or Mira-type variable), resulting in 32\,556 objects. The
positions of these stars were matched to the positions of objects in
the \WISE\ All-Sky source catalogue within 2 arcsec, rejecting sources
with more than one association. The resulting sample of 24\,942 stars
is further reduced by rejecting all sources with contamination and
confusion flags, as well as removing sources outside of saturation
limits (bands \w[1] through \w[3]) and 5$\sigma$ faint limits (see
Section~\ref{sec:intro}), yielding the final sample of 6171 objects
with presumably both reliable \WISE\ photometry and reliable SIMBAD
classifications.

We did not enforce any cut on Galactic coordinates. The distribution
of this sample in representative CC and CM diagrams constructed with
2MASS and \WISE\ photometry is shown in Fig.~\ref{fig:SIMBAD}. As
known from previous work, most AGB stars have 2MASS colour $J-K_s >
1$. Most of them also show clear \WISE\ colour excess, $\s[1,2] > 0$
and $\s[2,3] > 0$, as well as $K_s-\w[3]>0$. The two magnitudes limits
discernible from the source distribution in the top right panel are
due to adopted faint limit $\w[3] < 11.32$ and saturation limit $\w[1]
> 8.1$.

The latter is responsible for a significant difference between the
data distribution in our \c[2,3] versus \c[1,2] diagram (the bottom
left panel) and the diagram shown in fig.~4 from
\cite{TuWang2013}. Their version displays C stars below the dashed
line in our bottom left panel, and OH/IR stars above the dashed
line. Based on this separation, they proposed the dashed line as a
separator between these two classes of AGB stars. However, in our
version, all the stars below the line are absent, as well as all OH/IR
stars, and the region above the line is dominated by C stars, and not
by OH/IR stars! It turns out that these differences are caused by
rejecting stars brighter than saturation limits - when we relax this
criterion, we obtain a data distribution reminiscent of fig.~4 from
\cite{TuWang2013}. This is shown in Fig.~\ref{fig:saturated}, similar
to Fig.~\ref{fig:SIMBAD}, but now includes formally saturated
stars. All saturated stars, with \w[1]$<$8.1 and \w[2]$<$6.7, are
shown with red dots, while the black dots show unsaturated
sources. Notice in the lower left panel that almost all saturated
stars fall below the dashed line that \citet{TuWang2013} have devised
to separate C-rich from O-rich AGB stars. They did not mention
saturation limits in their work. We believe that sub-standard \WISE\
photometry for saturated AGB stars has induced unreliable source
distribution in fig.~4 from \cite{TuWang2013}. The \w[2] band is most
likely to blame, as evidenced by fig. 8 (in section VI.3.c.i.4) of the
\WISE\ Explanatory Supplement (see footnote~\ref{foot:explsuppl}),
where the bias for saturated sources is most pronounced in \w[2].

With saturation limits enforced, the distribution of C stars, by and
large from LMC and SMC, clearly stands out in Fig. \ref{fig:SIMBAD}
(see the bottom right panel). We define the 3D colour space
$x=\c[1,2]$, $y=\c[2,3]$, $z=\c[3,4]$. The box that outlines a high
concentration of C-rich AGB stars is defined by
\begin{align}
  \label{eq:N1}
  \phantom{-}1.59 x - 0.57 &< y < \phantom{-}1.59 x + 0.32\\
  \label{eq:N2}
            -0.63 x + 1.20 &< y <           -0.63 x + 5.07.
\end{align}

In the \c[2,3] versus \c[1,2] diagram shown in the bottom left panel
of Fig.~\ref{fig:SIMBAD}, O-rich stars with silicate dust are seen as
a low-density cloud towards the left from the C-rich star box. Further
insight in the distribution of O-rich and C-rich AGB stars in \WISE\
CC diagrams is based on complete \WISE\ samples and is discussed in
Section~\ref{sec:agb}.

Based on this analysis our samples of AGB stars for
Fig.~\ref{fig:agb-selection} are selected as follows. The initial
criteria are:
\[
\begin{array}{ll}
  \w[1]/\w[2]/\w[3]/\w[4] < 16.83/15.60/11.32/8.0 & 5\sigma \mbox{-limits}\\
  \w[1]/\w[2]/\w[3]/\w[4] > 8.1/6.7/3.8/{-}0.4  & \mbox{saturation limits}\\
  \w[1]<11,\ \w[2]<10 & \mbox{only bright sources}\\
  \mbox{SNR1 \& 2} > 5,\ \mbox{SNR3 \& 4} > 10 & \mbox{signal-to-noise ratio}
\end{array}
\]
The regions with $|b|<6$~deg were also excluded because YSOs tend to
cluster in the Galactic plane \citep{IE2000}, and also because of the
uncontrolled/unknown levels of Galactic extinction at these
latitudes. In the $xyz$ colour space defined above we also require $x
> 0.2$, $0 < y < 4$, excluding the long narrow plume of stars with
constant-density dust shells.  This so-selected initial AGB sample
consists of 8835 sources.

To select O-rich AGB stars from this initial sample we used the
colour-colour diagrams 3--5 in Fig.~\ref{fig:agb-selection} as a
guideline.  We defined a wedge-shaped volume bounded by these
criteria:
\begin{align}
\label{eq:O1}
 0.2 &< x < 2\\
 0.4 &< y < 2.2\\
   0 &< z < 1.3\\
   z &> 0.722 y - 0.289
\label{eq:O4}
\end{align}
The selected O-rich star sample has 1608 members.

The C-rich stars, visible as a strongly elongated plume in
Fig.~\ref{fig:agb-selection}, were selected by carving out a 3D
elongated volume around the plume. It is the intersection of the
following two slabs:
\begin{align}
\label{eq:C1}
0.629 y - 0.198 &< x < 0.629 y + 0.359\\
\label{eq:C2}
0.722 y - 0.911 &< z < 0.722 y - 0.289
\end{align}
Note that (\ref{eq:C1}) is the inverse of (\ref{eq:N1}), and that we
abandon (\ref{eq:N2}). The C-rich star sample contains 1383 sources.

For the C:O-rich star ratio map in Fig.~\ref{fig:co-ratio} we have
relaxed the $|b|<6$~deg requirement, and have selected C-rich and
O-rich star candidates through the CC volumes and magnitudes defined
above. The preselected sample contains 90\,114 sources, of which
14\,969 end up in the O-rich star group. The size of the C-rich star
selection grows to 1629 objects.

\section{LMC coordinate transformations}
\label{sec:lmcanalysis}

To obtain a de-projected map of the LMC disc, for our analysis of the
radial gradient of the C:O-rich star ratio in
Section~\ref{sec:co-ratio}, we follow the procedure outlined in two
papers on the structure of the Magellanic Clouds, \mci\ and \mcii. We
begin with the right ascension and declination angles $\alpha$ and
$\delta$ of our colour and magnitude selected C-rich and O-rich AGB
stars (given in degrees or radians). We further assume as the origin
of this equatorial coordinate system $(\alpha_0,\delta_o) =
(82\fdg25,-69\fdg5)$, as defined in Section 4.1 in \mci. This origin
corresponds to the centre of the density iso-contours of the
\emph{DENIS} survey, but is slightly offset w.r.t. the standard locus
of the LMC centre [e.g. $(\alpha_0,\delta_o) = (80\fdg9,-69\fdg76)$
given in the NED data base]. From equation~(1) of \mci\ we obtain the
formula for transformed angular coordinate $\rho$
\begin{equation}
  \label{eq:rho}
  \rho = \arccos\bigg\{\cos\delta\ \cos\delta_0\ \cos(\alpha-\alpha_0)\ +\ \sin\delta\ \sin\delta_0 \bigg\}.
\end{equation}
Merging equations~2 and 3 from the same paper we get the other
transformed coordinate $\phi$
\begin{equation}
  \label{eq:phi}
  \phi = \arctan\left\{ \frac{\sin\delta\ \cos\delta_0\ -\ \cos\delta\ \sin\delta_0\ \cos(\alpha-\alpha_0)}
                             {-\cos\delta\ \sin(\alpha-\alpha_0)} 
                \right\}.
\end{equation}
\mcii\ then defines in equations~2 the de-projected coordinates $x'$
and $y'$ as linear distances (in kpc) from the LMC centre, in the
plane of its disc
\begin{align}
  \label{eq:xp}
  x' &= \frac{D_0\ \cos i\ \sin\rho\ \cos(\phi-\Theta_{\rm far})} {\cos i\ \cos\rho\ -\ \sin i\ \sin\rho\ \sin(\phi-\Theta_{\rm far})},\\[5pt]
  \label{eq:yp}
  y' &= \frac{D_0\ \sin\rho\ \sin(\phi-\Theta_{\rm far})} {\cos i\ \cos\rho\ -\ \sin i\ \sin\rho\ \sin(\phi-\Theta_{\rm far})},
\end{align}
where $D_o$ is the distance from Earth to the LMC, 51~kpc, and
\hbox{$i=34\fdg7$} is the inclination of the LMC disc plane with
respect to our line of sight. $\Theta_{\rm far} = \Theta + 90^\circ$,
where $\Theta = 122\fdg5$ is the position angle (PA) of the ``line of
nodes'', the intersection of the LMC disc plane and the plane of the
sky. See the discussion in \mcii\ as to why $\Theta$ differs from the
PA of the LMC semi-major axis.

\mcii\ finally rotate the ($x'$,$y'$) coordinates by the angle
$\Theta_{\rm far}$ to align them with another set of coordinates in
their paper. For our work, this step is not necessary, but we execute
it to obtain a map oriented identically to \mcii\
\begin{align}
  \label{eq:xpp}
  x'' = x' \cos\Theta_{\rm far}\ -\ y' \sin\Theta_{\rm far},\\[5pt]
  y'' = x' \sin\Theta_{\rm far}\ +\ y' \cos\Theta_{\rm far}.
\end{align}
These are the coordinates we plot in the lower left panel of our
Fig.~\ref{fig:co-ratio-lmc}. We select all C-rich and O-rich stars
that fall within a $\sqrt{{x''}^2 + {y''}^2} \le 10$~kpc circle. Their
counts are, respectively, \hbox{$n_{\rm C} = 1114$} and \hbox{$n_{\rm
    O} = 252$}.

\section{Tabulation of model colour tracks and filter fluxes}
\label{sec:tabulation}

Equation~\eqref{eq:filterflux} averages the observed SED of a source
or model with the instrument's total transmission curves, and
calibrates the in-band fluxes to an empirical fit for Vega
\citep{Wright+2010}. Recall that the \WISE\ colour between two bands 1
and 2 is defined as
\begin{align}
  \c[1,2] &= 2.5\log\frac{f_2}{f_1} = 2.5 \left(\log \frac{F_2^{\rm src}}{F_1^{\rm src}} + \log \frac{F_1^{\rm Vega}}{F_2^{\rm Vega}} \right),\\
          &= (\c[1,2])^{\rm src} - (\c[1,2])^{\rm Vega}.
\end{align}
To compute colours, the absolute scale of the SED is irrelevant,
because all scales cancel out. However, to allow luminosity-dependent
studies using our models, we list in Table \ref{tab:tabulations},
together with the Vega-calibrated \WISE\ colours, the unnormalised
in-band fluxes $F_i^{\rm src}$ of the \D\ models used in
Figs~\ref{fig:compare-to-wise}, \ref{fig:D-colors}--\ref{fig:RDW}, and
\ref{fig:agb-analysis-cc}. The pure Vega in-band fluxes, using the
aforementioned empirical fit, are constants. Their values are:
\hbox{$F_i^{\rm Vega} = 5.576\times 10^{-13}, 2.956\times 10^{-13},
  7.289\times 10^{-14}, 4.360\times 10^{-15}$} for bands $i=$~1--4, in
the same units as the models. The \D\ model SEDs are normalized with
the bolometric flux, thus for comparison with observational data the
reported $F_i^{\rm src}$ values must be multiplied with the true
bolometric flux of the source.

The entire tabulations are available as an electronic supplement file
\texttt{model\_tables.dat} to this paper. All physical properties of
the models are listed in the header of each table, and are tabulated
in the electronic file. They are: the shell radial optical depth at
visual (0.55~\mic), the dust chemistry, grain size distribution, the
photospheric temperature $T_s$ of the central star, the radial density
distribution (either a power-law $r^{-p}$, or RDW), and the dust
temperature $T_d$ at the inner radius of the shell. All models have
radial shell thickness $Y = R_{\rm out}/R_{\rm in} = 100$.
\begin{table*}
 \begin{minipage}[b]{0.71\hsize}
  \begin{center}
    \caption{Tabulation (excerpt) of \WISE\ colours and
      filter-averaged fluxes $F_i^{\rm src}$ for all \D\ shell model
      tracks shown in Figures \ref{fig:compare-to-wise}, 
      \ref{fig:D-colors},                                
      \ref{fig:plume_yso},                               
      \ref{fig:yso_extended},                            
      \ref{fig:RDW}, and                                 
      \ref{fig:agb-analysis-cc},                         
      as a function of the shell radial optical depth at visual (left
      column). The entire tabulations are available as an electronic
      supplement to this paper (\texttt{model\_tables.dat}).}
    \label{tab:tabulations}
      \begin{tabular}{llccccccc}
        \hline
        \tV\      & \ldots & \c[1,2]   & \c[2,3]   & \c[3,4] & $F_1^{\rm src}$ & $F_2^{\rm src}$ & $F_3^{\rm src}$ & $F_4^{\rm src}$ \\
                  & \ldots & [3.4-4.6] & [4.6-12]  & [12-22] & & & & \\
        \hline
        \ldots    & \ldots & \multicolumn{7}{c}{\ldots} \\
        \multicolumn{9}{l}{\# Fig = 12, Ts = 14454 K, dust = ISM, grain\_size = MRN, Td = 600 K, density\_law = 1/r\^{}0.0} \\
        \# 0      & \ldots & 7         & 8         & 9         & 10        & 11        & 12        & 13 \\
        0.000e\p0 & \ldots & 5.780e\m3 & 1.541e\m2 & 1.303e\m2 & 2.102e\m3 & 1.114e\m3 & 2.640e\m4 & 1.779e\m5 \\
        1.000e\m4 & \ldots & 6.098e\m3 & 2.228e\m1 & 1.769e\p0 & 2.102e\m3 & 1.114e\m3 & 3.196e\m4 & 1.085e\m4 \\
        \ldots    & \ldots & \multicolumn{7}{c}{\ldots} \\
        \hline
      \end{tabular}
  \end{center}
 \end{minipage}
\end{table*}

\label{lastpage}

\end{document}